\documentclass[12pt]{article}
\usepackage{xcolor}
\usepackage{cite}
\usepackage{hyperref}
\usepackage{subcaption}
\usepackage[utf8]{inputenc}

\usepackage{setspace}
\usepackage{amsmath, amssymb, amsthm, float, graphicx}
\numberwithin{equation}{section}
\hypersetup{colorlinks=false, linkcolor=blue, citecolor=red}

\def\beq{\begin{eqnarray}}\def\eeq{\end{eqnarray}}
\def\be{\begin{equation}}\def\ee{\end{equation}}
\def\g{\gamma}

\def\s{\sigma}

\def\a{\alpha}
\def\e{\epsilon}

\def\b{\beta}
\def\d{\delta}
\def\c{\chi}

\def\D{\Delta}
\def\G{\Gamma}
\def\l{\lambda}

\def\Dphi{\Delta_{\phi}}

\def\G{\Gamma}

\def\dsdt{\frac{ds}{2\pi i}\frac{dt}{2\pi i}}
\begin{document}

\title{\bf On The Polyakov-Mellin Bootstrap}
\date{}

\author{\!\!\!\! Rajesh Gopakumar$^{a}$\footnote{rajesh.gopakumar@icts.res.in}~and Aninda Sinha$^{b}$\footnote{asinha@iisc.ac.in}\\ ~~~~\\
\it ${^a}$International Centre for Theoretical Sciences (ICTS-TIFR),\\
\it Shivakote, Hesaraghatta Hobli,
Bangalore North, India 560 089\\
\it ${^b}$Centre for High Energy Physics,
\it Indian Institute of Science,\\ \it C.V. Raman Avenue, Bangalore 560012, India. }
\maketitle
%\author{Kallol Sen\footnote{kallol@cts.iisc.ernet.in} ~and Aninda Sinha\footnote{asinha@cts.iisc.ernet.in}\\ ~~~~\\
%\it Centre for High Energy Physics,
%\it Indian Institute of Science,\\ \it C.V. Raman Avenue, Bangalore 560012, India. \\}
\maketitle
\vskip 2cm
\abstract{We elaborate on some general aspects of the crossing symmetric approach of Polyakov to the conformal bootstrap, as recently formulated in Mellin space. This approach uses, as building blocks, Witten diagrams in AdS.  We show the necessity for having contact Witten diagrams, in addition to the exchange ones, in two different contexts: a) the large $c$ expansion of the holographic bootstrap b) in the $\epsilon$ expansion at subleading orders to the ones studied already. In doing so, we use alternate simplified representations of the Witten diagrams in Mellin space. This enables us to also obtain compact, explicit expressions (in terms of a ${}_7F_6$ hypergeometric  function!) for the analogue of the crossing kernel for Witten diagrams i.e., the decomposition into $s$-channel partial waves of crossed channel exchange diagrams. 
%Other technical spinoffs include include explicit expressions for the crossing kernel, which expresses the $t$-channel exchange diagram in terms of s-channel double trace blocks.}

\tableofcontents
%\tableofcontents

\onehalfspacing

\section{Introduction}
Since the resurgence of interest in the conformal bootstrap in spacetime dimensions greater than two \cite{rrtv}, there have been considerable advances in developing both numerical and analytical tools to determine the conformal data i.e., the spectrum of dimensions and OPE coefficients. See, for instance, the recent review \cite{prv}. Typically the approach has been to impose crossing symmetry on the four point function leading to bootstrap equations which relate an expansion of the $s$-channel amplitude, in conformal blocks, with the $t$-channel one \cite{stuff,alday1,aldayeps,azz, aldayloops,holorecon, fkps,kz,dsd, kss}. Here each channel has only the physical operators appearing in the OPE of two of the external operators. Crossing symmetry then imposes nontrivial constraints on the conformal data. 

Recently, an alternative approach proposed by Polyakov in 1974 \cite{polya}  was revived in a modern incarnation \cite{usprl,longpap,dks,dgs} while also gaining in technical  power through adopting the machinery of Mellin space \cite{mack,mellinjoao,fitz,pau, regge, spinningads, othermelpaps}(see \cite{KSAS} for an earlier attempt in momentum space). 
Polyakov proposed expanding the four point function of a $CFT_d$ in terms of manifestly crossing symmetric building blocks and then imposing consistency with the OPE as a constraint. Rather remarkably, the crossing symmetric blocks suggested by Polyakov can now be interpreted as essentially exchange  Witten diagrams in $AdS_{d+1}$ \cite{usprl, longpap}. When we expand a four point function in terms of these blocks, we generally will also have  contributions from what are called ``double-trace" operators in AdS/CFT. In the case of identical external scalars (with dimension $\Dphi$), these are operators with dimension $\D_{n,\ell}=2\Dphi+2n+\ell$. In other words, we make an expansion of the form
\be\label{PMexp}
{\cal A}(u,v)=\sum_{\Delta,\ell} c_{\Delta,\ell} \big(W^{(s)}_{\Delta,\ell}(u,v) +W^{(t)}_{\Delta,\ell}(u,v)+W^{(u)}_{\Delta,\ell}(u,v)\big) \,,
\ee
where $W^{(s)}_{\Delta,\ell}(u,v)$ is the $s$-channel exchange diagram corresponding to an operator
of dimension $\D$ and spin $\ell$ in the $CFT_d$ and similarly for the other channels. These are conformally invariant by construction and thus have an expansion in terms of ($s$-channel) conformal blocks which is schematically of the form of the (nontrivial part) of the four point function
\be\label{wittconf}
W^{(s)}_{\Delta,\ell}(u,v) \sim G^{(s)}_{\Delta,\ell}(u,v)  +\sum_{n,\ell} \big(a^{\D, \ell}_{n,\ell'} G^{(s)}_{\Delta_{n,\ell'},\ell'}(u,v) +b^{\D, \ell}_{n,\ell'} \partial_{\D}G^{(s)}_{\Delta_{n,\ell'},\ell'}(u,v) \big) \, .
\ee
For the crossed channels, we have only a sum over the double trace operators i.e., without the first term in eq.(\ref{wittconf}). 
  
In a large $N$ CFT there are physical primaries in the spectrum whose dimensions are approximately $\D_{n,\ell}$ upto $\frac{1}{N}$ corrections, and the additional contributions in the Witten exchange diagram are conventionally interpreted as leading terms in a $\frac{1}{N}$ expansion of the anomalous dimensions. However, in an arbitrary CFT, if we view the exchange diagrams as a convenient set of functions to expand in, there is typically\footnote{Exceptions would be either free theories or subsectors of protected operators in a SUSY theory, which can be dealt with separately.} no reason to have physical operators with dimension exactly given by $\D_{n,\ell}$. Thus we view these as giving rise to spurious power law/logarithmic dependence on the cross ratios such as $u^{\Dphi+n}, u^{\Dphi+n}\log{u}$, or equivalently spurious single and double poles in the mellin variable $s$ (which is conjugate to the cross ratio $u$). The OPE  consistency condition, in this Polyakov-Mellin approach to the bootstrap, is to impose the vanishing of the residues of these spurious poles once one sums over the full spectrum. The Mellin space analogue of eq.(\ref{PMexp}) is obtained by the replacements 
(see beginning of Sec. 2 for precise definitions)  
\be
{\cal A}(u,v) \rightarrow {\cal M}(s,t); \, \, \, \,W_{\Delta,\ell}^{(s)}(u,v) 
\rightarrow M_{\Delta,\ell}^{(s)}(s,t) \, .  
\ee

When we look at the residue at the spurious poles, $s=\Dphi+r$, the issue arises about the convergence\footnote{The convergence issue in the conventional approach has been addressed in \cite{rych}.} of the sum over the spectrum for different values of $t$. As we will see, we will be expanding the residues in terms of a set of orthogonal partial wave polynomials in $t$. The validity of this expansion requires that the sum converge (in each channel) in some neighbourhood of $t=0$. One way to argue that this is possible is as follows. In the $t$ (and $u$)-channel we have physical poles corresponding to exchanged operators which would be seen in the $s$-channel only after summing over the spectrum. The poles in the former are at $t=\frac{\D-\ell}{2}-\Dphi+q$ and at $t=-s+ \Dphi- \frac{\D-\ell}{2}-m$. We therefore see that convergence is possible only if we are within the closest pole. Thus imposing the  most stringent inequalities 
\be\label{tpole}
|t | \leq \Big|\frac{d-2}{2} -\Dphi \Big|\,.
\ee
Here we have used the unitary bounds on the twist $\tau =\D-\ell \geq d-2$ for $\ell > 0$. 
For $\ell =0$ we have $\tau > \frac{d-2}{2}$ and hence a bound $|t | \leq \Big| \frac{d-2}{4} -\Dphi  \Big |$. 
%{(\bf RG: Note this is somewhat different from that stated in your email.)}

It was shown in \cite{usprl, longpap} that this approach can successfully be implemented in situations like the $\epsilon$ expansion of the Wilson-Fisher fixed point. It allowed one to obtain not only anomalous dimensions up to $O(\e^3)$ for $\Dphi$ and twist two operators (for $\ell=0$ and twist two, the anomalous dimension was obtained correctly up to $O(\e^2)$) but also the corresponding OPE coefficients. The latter are quantities  not yet computed, in general, even with Feynman diagrams. These results have since also been generalised to the case with $O(N)$ symmetry \cite{dks}, for a leading order perturbative proof for non-existence of CFTs beyond 6 dimensions \cite{golden} as well as to study the epsilon expansion in the large spin limit \cite{dk}. 

In our earlier work, we used only the exchange Witten diagrams as crossing symmetric building blocks and did not explicitly add in contact diagrams which are also crossing symmetric. Indeed they did not seem necessary to reproduce the results of the $\e$ expansion, at least to the order studied. However, it is somewhat unnatural to leave out contact Witten diagrams. In fact, there are ambiguities, in how one defines exchange Witten diagrams, coming from the choice of the cubic vertex. As we will see, some of these ambiguities can be traded for contact Witten diagrams. Hence it is not clear there is a physically invariant notion which allows one to drop the contact diagrams. At the same time, if we do add the contact diagrams, the question arises as to what determines their coefficients. Note that for the exchange diagrams, the coefficients $c_{\D,\ell}$ in eq. (\ref{PMexp}) were essentially proportional to the square of the three point OPE coefficients $C_{\phi\phi\cal{O}}$, which is part of the conformal data one is seeking to solve for. In this paper, while we will not be able to nail down the criteria which fixes the contact terms, we will nevertheless make some progress in addressing the issue of their presence. 

We will discuss the appearance of contact terms in two different situations. The first is in the holographic bootstrap where we revisit the work of Heemskerk et.al \cite{heemsk}. In a simplified toy version of AdS/CFT they showed that the solutions of the CFT crossing equations, in a large $N$ or generalised free field (GFF) expansion, are in one to one correspondence with bulk AdS vertices. We show the connection between these two pictures and our third picture which, despite being framed in terms of AdS objects, is different from the other two. We show how we recover the results of \cite{heemsk} by including contact Witten diagrams. Indeed we find it is inconsistent not to do so thus indicating they might necessarily have to be there, in general. We also look at the $\e$ expansion where we find that without contact diagrams we would not be able to reproduce the correct results when going to higher orders in $\e$.  Finally, we show how we can actually obtain additional new results, for some leading twist four operators, in the $\e$ expansion using our present approach to the bootstrap. 
  
 In obtaining these results, we lean heavily on some substantial technical simplifications in the machinery that we employ. In the formalism of \cite{usprl,longpap}, the actual calculations, while being conceptually straightforward, were technically involved and the expressions were not in the most manageable form for general calculations. Here we are able to essentially overcome these technical challenges in the following way
\begin{itemize}
\item
Firstly, for the Witten exchange diagrams in Mellin space we will simplify the spectral representation
that was used in \cite{usprl, longpap} and work with a meromorphic form which is given explicitly in terms of the Mack Polynomials and a ${}_3F_2$ hypergeometric function (see eq.(\ref{meromwitt})). 
We exploit the ambiguity in adding contact diagrams to fix this form.
\item
We parametrise the possible contact Witten diagrams in a natural way, following \cite{heemsk}, which allows one to see exactly what the ambiguities are. 
\item
Finally, and perhaps most nontrivially, we use the meromorphic form of the Witten diagram mentioned above to explicitly compute the analogue of the crossing kernel for the Witten diagrams. Recall that the usual crossing kernel refers to an expansion of the $t$-channel conformal block in terms of $s$-channel blocks. Here we are considering the analogous expansion of the cross channel Witten exchange diagrams such as $M^{(t)}_{\Delta,\ell}(s,t)$ in terms of the $s$-channel (double-trace) conformal blocks. This is important to get an explicit form for the equation for the vanishing residues at spurious poles, which imposes consistency in our approach. We give a complete solution in the case of identical external scalars but for arbitrary intermediate exchanged operators. This has a remarkably compact form in terms of a finite sum of ${}_7F_6$ hypergeometric functions. These latter are closely tied to the $6j$-symbols for the conformal group that have also been investigated recently \cite{gadde, vanrees, tar-slght, 6jdsd} i.e., the usual crossing kernel (see also \cite{Cardona:2018dov, tar-slght2}). 
\end{itemize} 

%However, in terms of Witten diagrams in AdS, there are also contact diagrams. It was not clear if such diagrams should also be added to the basis and if so what the evidence for this is. One of the goals of this paper will be to clarify the situation regarding this.....

%Furthermore, in the formalism of \cite{usprl,longpap,PDAKAS}, the actual calculations, while being conceptually straightforward, were technically involved. The main reason for this complication was the use of a spectral function. As a result general expressions for the constraint conditions were hard to find. Our goal in this paper is to overcome this shortcoming and present a technically simpler approach.

%subsection{A different perspective on crossing symmetric blocks}

In this context, let us make a comment on how
employing the meromorphic form for the Witten diagrams helps to give a perspective on their use in the crossing symmetric blocks -- it suggests why Polyakov was naturally led to these as his building blocks. As we will see, the $s$-channel exchange diagram in Mellin space reads as
\be\label{meromexp}
M^{(s)}_{\Delta,\ell}(s,t)=\sum_q \frac{f_q(s,t)}{s-\frac{\Delta-\ell}{2}-q}\, ,
\ee
with $f_q(s,t)$ being polynomials. 
As discussed above, the Witten diagram expansion is designed  so that, after the spurious pole cancellation, it coincides with the usual $s$-channel (or for that matter, crossed channel) expansion. 
The double trace contributions then cancel out and we are left with only the physical operator contributions. We are able to do this since the meromorphic form of the Witten exchange diagram in eq.(\ref{meromexp}) has the same poles and residues as the conformal block. Therefore we can alternatively rewrite the usual $s$-channel conformal block expansion of the amplitude as 
\be\label{usual}
{\cal A}(u,v) = \sum_{\Delta,\ell} c_{\Delta,\ell} \int \dsdt\, u^s v^t M^{(s)}_{\Delta,\ell}(s,t) \frac{\sin^2 \pi(\Delta_\phi-s)}{\sin^2 \pi(\Delta_\phi-\frac{\Delta - \ell}{2})}\rho_{\Dphi}(s,t)\,,
\ee
where we have replaced the conformal block with $M^{(s)}_{\Delta,\ell}(s,t)$ together with a $\sin^2 \pi(\Delta_\phi-s)$ factor.  With the additional $\sin^2 \pi(\Delta_\phi-s)$ factor we get rid of the double poles from the measure $\rho_{\Dphi}(s,t)=\Gamma^2(\D_\phi-s)\Gamma^2(s+t)\Gamma^2(-t)$ which correspond to the (spurious) double trace operators. 

One can thus motivate crossing symmetric blocks as follows. Due to the $\sin^2 \pi(\Delta_\phi-s)$ factor, the integrand in the Mellin representation of the s-channel block behaves like $e^{2\pi |s|}$ for large imaginary $s$ (since $M^{(s)}_{\Delta,\ell}(s,t)$ is polynomially bounded at large $s,t$)--see also \cite{fkap}. Thus with the plain conformal blocks we have a bad behaviour at infinity due to this factor and hence these are not the best set of functions to expand in a crossing symmetric way. However, if we get rid of this factor, the penalty to be paid is to introduce spurious poles that come from the measure factor. As mentioned above, these will not be consistent with the OPE as there are no operators whose dimensions are precisely $2\D_\phi+2n$. The only chance for the expansion to be consistent is to add the crossed channel so as to have explicit crossing symmetry and then demand that the spurious poles cancel. This entails using 
\be\label{witten}
{\cal A}(u,v)=   \int \dsdt\, u^s v^t \Big(\sum_{\Delta,\ell} c_{\Delta,\ell} \big(M^{(s)}_{\Delta,\ell}(s,t) +M^{(t)}_{\Delta,\ell}(s,t)+M^{(u)}_{\Delta,\ell}(s,t)\big)+M^{(c)}(s,t)\Big)\rho_{\Dphi}(s,t)\,,
\ee
where $M^{(c)}(s,t)$ is a manifestly crossing symmetric regular (possibly polynomial) function. These are presumably parametrised by contact Witten diagrams. 
%An efficient way of parametrizing the crossing symmetric polynomial is:
%\be
%d(s,t)=a_0+\sum_{\{m,n\}} a_{mn} \left[s(s+t-\D_\phi)(t+\D_\phi)\right]^m \left[t(s+t)+s(s-\D_\phi)\right]^n\,,
%\ee
%where $a_{mn}$'s are constants .

In the next section, starting from the spectral (or `split') representation of the exchange Witten Diagram, we derive the meromorphic form which is what we will employ later. We also usefully parametrise the contact Witten diagrams in increasing polynomial degree. Sec. 3 uses the results of Sec. 2 to give a complete solution to the contribution of any given exchange diagram to the spurious residues of a particular partial $s$-wave (for identical scalars). While this is relatively straightforward for the $s$-channel exchange diagram (Sec. 3.1), it is complicated for the crossed channels. We study some nice special cases in Sec. 3.2 and give explicit expressions for the general case in Sec. 3.3 and appendix \ref{appd}. In Sec. 4 we come to the case of the holographic bootstrap where we argue that the crossing symmetric approach with contact diagrams precisely reproduces the leading order results. 
Sec. 5 revisits the $\e$ expansion and discusses the role of contact diagrams here and argues that they are needed when we go to higher orders in $\e$. A couple of new results for twist four operators derived using bootstrap are also recorded. We conclude with a discussion section while sequestering some of the more technical details in various appendices. 

\section{Witten Diagrams in Mellin Space}

This section and the next are somewhat technical. In this section we describe a couple of different useful forms in which we can cast Witten exchange diagrams in Mellin space (with identical external scalars for simplicity). These results will play a direct role in our considerations of the following sections. We will also systematise the discussion of the contact Witten diagrams in Mellin space. However, the reader who is not into Mellin minutiae can skip 
this section from which we will mainly use Eqs.(\ref{meromwitt}) and (\ref{wittcont}). 

We employ the same notation as in \cite{longpap}.
In particular, the Mellin representation of the nontrivial part of the four point correlator ${\cal A}(u,v)$ is defined as 
\be\label{idmelldef}
{\cal A}(u,v)= \int_{-i\infty}^{i\infty}\frac{ds}{2\pi i} \, \frac{dt}{2\pi i} \, u^{s}v^t \Gamma^2(-t)\Gamma^2(s+t)\Gamma^2(\Delta_{\phi}-s){\cal M}(s,t)\,.
\ee
The measure factor (due to Mack) $\Gamma^2(-t)\Gamma^2(s+t)\Gamma^2(\Delta_{\phi}-s)$ will be denoted as 
$\rho_{\Delta_{\phi}}(s,t)$.  
Under this transform, various position space entities have their corresponding Mellin space representation.
Thus for the Witten exchange diagrams, labelled by the dimension ($\Delta$) and spin ($\ell$) of the exchanged operator (with identical external scalars as mentioned above) 
\be
W_{\Delta,\ell}^{(s)}(u,v) \rightarrow M_{\Delta,\ell}^{(s)}(s,t).  
\ee
The superscript denotes the channel (in this case $s$) and should not be confused with the Mellin variables. 
In the following we will describe various useful forms for $M_{\Delta,\ell}^{(s)}(s,t)$. 

\subsection{Spectral Representation of Exchange Diagrams}

The Witten exchange diagram being conformally invariant, can be decomposed in terms of the so-called conformal partial waves $F^{(s)}_{\nu,\ell}(u,v)$ which are linear combinations of the corresponding conformal blocks and their shadow (see for example Eq.(2.15) of \cite{longpap}). These are a set of orthonormal functions labelled by the principal series representations of the conformal group (for which the parameter $\nu$ is purely imaginary). In Mellin space
\cite{dolanosborn2}
\begin{align}
\begin{split}\label{sparwavemell}
F_{\nu,\ell}^{(s)}(u,v)=& \int_{-i\infty}^{+i\infty} \frac{ds}{2\pi i}\ \frac{dt}{2\pi i}\ u^s v^t \rho_{\D_\phi}(s,t)\Omega_{\nu, \ell}^{(s)}(s)P^{(s)}_{\nu, \ell}(s,t). 
\end{split}
\end{align} 
Here $P^{(s)}_{\nu, \ell}(s,t)$ are the so-called Mack polynomials (of degree $\ell$ in $(s,t)$) which  are explicitly given in appendix \ref{B} and  
\be\label{omdef}
\Omega_{\nu, \ell}^{(s)}(s)=\frac{ \G(\frac{h+\nu-\ell}{2}-s)\G(\frac{h-\nu-\ell}{2}-s)}{\G^2(\Delta_\phi-s)}. 
\ee
The Mellin exchange amplitude $M_{\Delta,\ell}^{(s)}(s,t)$ can be written as
\be\label{wspecdecomp}
M_{\Delta,\ell}^{(s)}(s,t) = \sum_{\ell^\prime =0}^{\ell} \int_{-i\infty}^{+i\infty} d\nu \mu^{(s)}_{\Delta, \ell'}(\nu)
\Omega_{\nu, \ell'}^{(s)}(s)P^{(s)}_{\nu, \ell'}(s,t).
\ee
The spectral density $\mu^{(s)}_{\Delta, \ell'}(\nu)$ has poles in $\nu$ such that when the contour integral over 
$\nu$ is done, by closing on the right half plane, we pick up their contributions and this tells us which conformal blocks appear in the amplitude. All the physical information is, in fact, captured by the term with $\ell'=\ell$. 
\be\label{specunitry0}
\mu^{(s)}_{\Delta,\ell}(\nu)=\frac{\G^2(\D_\phi -\frac{h-\ell-\nu}{2})\G^2(\D_\phi - \frac{h-\ell+\nu}{2})}{2\pi i ((\D-h)^2-\nu^2)\G(\nu)\G(-\nu)(h+\nu-1)_\ell (h-\nu-1)_\ell } \, .
\ee
We see here that there is a pole corresponding to the physical operator that is exchanged, at $\D=(h+\nu)$ (as well as the shadow pole which is not picked up). In addition, the gamma functions in the numerator give rise to a series of double trace poles which correspond in position space to conformal blocks (as well as their derivatives with respect to the dimension) of operators of spin $\ell$ and twist $\tau = 2\D_\phi+2n$. However, we see from the form of Eq.(\ref{omdef}) that these poles do not actually show up as poles in the $s$-variable after doing the $\nu$ integral. Instead, the double trace operators show up in the $s$-variable through the dependence in the measure term $\rho_{\Delta_{\phi}}(s,t)$.  

There are additional poles from the Pochammer terms in the denominator: $(a)_m \equiv \frac{\G(a+m)}{\G(a)}$. These are cancelled by similar poles from the other terms $\mu^{(s)}_{\Delta, \ell'}(\nu)$ with $\ell' < \ell$. We will not directly need these lower order terms. The form of these terms are essentially fixed (see \cite{spinningads}, Eq. (152) and Appendix E, therein) except for the constant pieces which we discuss later.   

One of the advantages of the spectral representation is that on carrying out the $\nu$-contour integral the Mack polynomials $P^{(s)}_{\nu, \ell}(s,t)$ are evaluated at the $\nu$-values corresponding to the poles. We can then use the nice property of Mack polynomials that
\be
\frac{4^\ell }{(\Delta-1)_{\ell}(2h-\Delta-1)_{\ell}}P^{(s)}_{\Delta-h,\ell}(s=\frac{\Delta-\ell}{2}+m,t)  \equiv Q^{\Delta}_{\ell,m}(t). 
\ee
These polynomials on the RHS (of degree $\ell$ in the $t$-variable) have many nice properties. In particular, for $m=0$, the $Q^{\Delta}_{\ell,0}(t)$ are a set of orthogonal polynomials - the continuous Hahn polynomials - which play the role that the Legendre or Gegenbauer polynomials do for spherical symmetry. We write down their explicit form and their orthogonality properties in Appendix \ref{A}

This was for the $s$-channel exchange. The $t, u$-channel exchange amplitudes in Mellin space can be similarly written down by appropriate exchange (with shifts) of the Mellin variables in the above expressions. See sec. 2.2 of \cite{longpap}. We will write them down explicitly in the next section in the alternate meromorphic form.

\subsection{Meromorphic Form of the Exchange Diagrams}

In \cite{longpap} we had taken the {\it definition} of the Witten diagram to be what one gets from the physical $\nu$ pole contributions on doing the contour integral in the spectral representation of Eq.(\ref{wspecdecomp}). This involved a minimal prescription of only focussing on the 
$\ell'=\ell$ term in the spectral representation Eq.(\ref{wspecdecomp}), since that contains all the physical poles, and ignoring the rest of the contributions since their role is only to cancel out the additional unphysical poles. While this is fine as a prescription it should be recognised that implicit in this prescription is a choice of terms which are entire in $(s,t)$. In effect we are fixing an ambiguity in the choice of our exchange diagram which is actually a polynomial of degree $\ell$ in ($s,t$)\footnote{The entire terms come from the leftovers after cancelling out the additional $\nu$-pole contributions which give unphysical poles in $s$. These come from evaluating the prefactors of $\Omega_{\nu, \ell'}^{(s)}(s)P^{(s)}_{\nu, \ell'}(s,t)$ at the additional $\nu$ poles and picking out the pieces which are entire in $(s,t)$. The polynomial dependence on $(s,t)$ follows from the form of the expressions for $\mu^{(s)}_{\Delta, \ell'}(\nu)$ given in \cite{spinningads}, from the $\nu$ dependence of the Mack Polynomial also being polynomial and from $\Omega_{\nu, \ell'}^{(s)}(s) =P(s)\Omega_{\nu, \ell}^{(s)}(s)$ with $P(s)$ a polynomial for $\ell' < \ell$.}.
  
The spectral representation gives a transparent way of exhibiting the conformal decomposition of the Witten diagrams. However, it is often cumbersome to work with while implementing the Polyakov-Mellin bootstrap idea. We will find it easier to use a somewhat different representation which more directly exhibits the analytic structure in $(s,t)$ of the Witten diagram in Mellin space. 
In particular, as has been noted in the literature on the Mellin representation, the Witten diagrams are meromorphic in $(s,t)$ with the residues of the poles being exactly that of the corresponding conformal blocks in Mellin space. However, unlike the conformal blocks, which behave exponentially in $(s,t)$ at infinity, the Witten exchange diagrams grow at most polynomially (of degree $(\ell -1)$, for an exchanged field of spin $\ell$), as was observed in the introduction as well.  

In this section, we will treat the $\nu$ integral somewhat differently, again focussing on the terms which contribute to the physical poles in $s$. This will give us a simple closed form expression which captures the meromorphic piece of the exchange contribution but will potentially differ from that of the previous section by terms which are polynomial in $(s,t)$. In other words, this is a {\it different} choice in fixing the polynomial ambiguity of the exchange diagram. 

We start again with the $\ell'=\ell$ term of the spectral function integral Eq.(\ref{wspecdecomp}). 
Denoting for simplicity
$$
\widehat P^{(s)}_{\nu,\ell}(s,t)=\frac{ P^{(s)}_{\nu,\ell}(s,t)}{(h+\nu-1)_\ell (h-\nu-1)_\ell}
$$  
we decompose the piece 
\be\label{mackdecomp}
\frac{\widehat P^{(s)}_{\nu,\ell}(s,t)}{(\Delta-h)^2-\nu^2}=\frac{\widehat P^{(s)}_{\Delta-h,\ell}(s,t)}{(\Delta-h)^2-\nu^2} + F_{up}(\nu, s,t)+\tilde{P}_c(\nu, s,t)\, .
\ee
Here we have separated out the first term which corresponds to the physical pole for the exchanged operator. The remaining terms are clubbed as follows. The ones which have additional poles in $\nu$ (coming from the pochammer piece in the denominator of Eq.(\ref{specunitry0}) ) give rise to unphysical poles in $s$. These are denoted as $F_{up}(\nu, s,t)$. They will cancel out after taking into account all the additional terms with $\ell' < \ell$. 
The last term of $\tilde{P}_c(\nu, s,t)$ denotes the terms which, after carrying out the $\nu$ integral, are polynomial in $(s,t)$. Note that there is an ambiguity in this separation into $F_{up}(\nu, s,t)$ and $\tilde{P}_c(\nu, s,t)$ since different choices in the former can differ by terms which are polynomial. 

This motivates one to make an alternative prescription for the exchange Witten diagram in which we only consider the first term in Eq.(\ref{mackdecomp}) while carrying out the $\nu$ integral. Again this will capture the contribution of physical poles in $s$. But it differs from the previous prescription by terms which are polynomial in $(s,t)$. One can see this from the fact that the Mack polynomials which appear in the present prescription always are of the form $P^{(s)}_{\Delta-h,\ell}(s,t)$ (i.e. with $\nu=\Delta-h$). Whereas in the prescription of Sec. 2.1 of doing the 
$\nu$ contour integral we would have Mack polynomials with other values of $\nu$. We will explicitly illustrate the difference for the first nontrivial case of $\ell =2$ in appendix \ref{C}.  

Replacing the LHS of Eq.(\ref{mackdecomp}) with the first term of the RHS in the $\ell'=\ell$ term of Eq.(\ref{wspecdecomp}), we are left with the integral 
\be\label{meromint}
M_{\Delta,\ell}^{(s)}(s,t) =  {\widehat P^{(s)}_{\Delta-h, \ell}(s,t)} \int_{-i\infty}^{+i\infty} \frac{d\nu}{ 2\pi i} 
\frac{\G^2(\D_\phi -\frac{h-\ell-\nu}{2})\G^2(\D_\phi - \frac{h-\ell+\nu}{2})}{((\D-h)^2-\nu^2)\G(\nu)\G(-\nu)}\frac{ \G(\frac{h+\nu-\ell}{2}-s)\G(\frac{h-\nu-\ell}{2}-s)}{\G^2(\Delta_\phi-s)} \, .
\ee

This $\nu$ integral can be carried out using the useful identity.
\begin{eqnarray}\label{id1}
\int_{-i\infty}^{i\infty} \frac{\ d\nu}{2\pi i} \frac{\displaystyle\prod_{i=1}^3\Gamma(a_i-\frac{\nu}{2})\Gamma(a_i+\frac{\nu}{2})}{(4a_4^2-\nu^2)\Gamma(\nu)\Gamma(-\nu)}&=&\frac{\Gamma(a_1+a_2)\Gamma(a_1+a_3)\Gamma(a_2+a_4)\Gamma(a_3+a_4)}{(a_1+a_4)\Gamma(1+2a_4)}\nonumber \\ &\times& {}_3F_2[a_1+a_4,1-a_2+a_4,1-a_3+a_4;1+a_1+a_4,1+2a_4;1]\nonumber \,.\\
\end{eqnarray}
This identity has been implicitly used in \cite{mellinjoao, pau} without proof. It, however, does not appear to be a well known identity in the literature\footnote{A mathematica notebook proving this can be provided on request.}. Note that the manifest symmetry between the $a_i$ on the LHS is not apparent on the RHS but is nevertheless true using the transformation properties of the $_3F_2$. We take $a_1=\frac{h-\ell}{2}-s$ and find
for Eq.(\ref{meromint})
\begin{eqnarray}\label{meromwitt}
M_{\Delta,\ell}^{(s)}(s,t) &=& {\widehat P^{(s)}_{\Delta-h, \ell}(s,t)} \frac{\Gamma^2 \left(\frac{\Delta+\ell }{2}+\Delta_\phi -h\right)}{(\frac{\Delta-\ell}{2}-s) \Gamma (\Delta-h +1)} \,  \\
&\times & _3F_2\left[\frac{\Delta -\ell}{2}-s,1+\frac{\Delta-\ell }{2}-\Delta_\phi ,1+\frac{\Delta-\ell }{2}-\Delta_\phi ;1+\frac{\Delta-\ell }{2}-s,\Delta-h +1;1\right]\,.\nonumber
\end{eqnarray}
This is the form that we will use from now on. As mentioned above, we have moved the polynomial ambiguities to contact diagrams which will be discussed in the next subsection.

Note that in terms of the usual Mack Polynomials (as used for e.g. in \cite{longpap})
$$
{\widehat P^{(s)}_{\Delta-h, \ell}(s,t)}=P^{(s)}_{\Delta-h, \ell}(s,t)\frac{\Gamma(\Delta-1)\Gamma(2h-\Delta-1)}
{\Gamma(\Delta +\ell-1)\Gamma(2h-\Delta+\ell-1)} \, .
$$
We also note that the $_3F_2[a_i;b_j;1]$ function is well defined only for $b_1+b_2-(a_1+a_2+a_3)>0$. In Eq.(\ref{meromwitt}) this translates to the condition $2\Delta_\phi-h+\ell >0$. In the $\e$ expansion this does not hold  for the scalar channel ($\ell =0$). In such cases we can use the analytic continuation of the  $_3F_2$ in its parameters 
\be\label{3f2t}
_3F_2[a_1,a_2,a_3;b_1,b_2;1]=\frac{\Gamma(b_1)\Gamma(b_1+b_2-a_1-a_2-a_3)}{\Gamma(b_1-a_1)\Gamma(b_1+b_2-a_2-a_3)}{}_3F_2[a_1,b_2-a_2,b_2-a_3;b_2,b_1+b_2-a_2-a_3;1] \, .
\ee

Proceeding for the moment with the expression in Eq.(\ref{meromwitt}), we use the series expansion of the 
$_3F_2$
\begin{eqnarray}\label{3f2pole}
& &\frac{1}{\Gamma (\Delta-h +1)} \, _3F_2\left[\frac{\Delta -\ell}{2}-s,1+\frac{\Delta-\ell }{2}-\Delta_\phi ,1+\frac{\Delta-\ell }{2}-\Delta_\phi ;1+\frac{\Delta-\ell }{2}-s,\Delta-h +1;1\right] \nonumber \\ &=& \sum_{r=0}^\infty \frac{(1+\frac{\Delta-\ell }{2}-\Delta_\phi)_r^2(\frac{\Delta-\ell}{2}-s)}{r! \Gamma(\Delta-h +1+r)}\frac{1}{\frac{\Delta -\ell}{2}-s+r}, 
\end{eqnarray}
and see that the present prescription for the Witten exchange diagram is of a simple meromorphic form 
\be\label{meromwitt2}
M_{\Delta,\ell}^{(s)}(s,t) =  \Gamma^2 \Big(\frac{\Delta+\ell }{2}+\Delta_\phi -h\Big){\widehat P^{(s)}_{\Delta-h, \ell}(s,t)} \sum_{n=0}^\infty \frac{(1+\frac{\Delta-\ell }{2}-\Delta_\phi)_n^2}{n! \Gamma(\Delta-h +1+n)}\frac{1}{\frac{\Delta -\ell}{2}-s+n} \, .
\ee
At each of the poles the residue is proportional to 
$$
P^{(s)}_{\Delta-h, \ell}(s,t)|_{s=\frac{\Delta -\ell}{2}+n} \propto Q^{\Delta}_{\ell,n}(t),
$$
as expected. The dependence on $t$ is always a bounded polynomial (of degree $\ell$). 

In the case of a scalar with $\ell=0$, the prescription is particularly simple since the Mack Polynomial is just a constant. 
\be
M_{\Delta,\ell=0}^{(s)}(s,t) \propto \, _3F_2\left[\frac{\Delta}{2}-s,1+\frac{\Delta}{2}-\Delta_\phi ,1+\frac{\Delta}{2}-\Delta_\phi ;1+\frac{\Delta}{2}-s,\Delta-h +1;1\right]\, ,
\ee
with simple poles having constant residues. Note there is no $t$-dependence whatsoever. 

\subsection{Contact Diagrams in Mellin Space}

We have been considering the exchange Witten diagrams thus far. As we saw, different prescriptions for these diagrams differ by polynomials in the Mellin variables $(s,t)$. These correspond to the contact Witten diagrams. 
Since we have this ambiguity, we must allow for the possibility of adding arbitrary contact Witten diagrams while expanding a four point function.  Contact diagrams can be taken to be crossing symmetric in all channels when the external scalars are identical. Crossing symmetry acts, in our conventions, on the Mellin variables as 
\be\label{channexch}
S-T \, {\rm exchange}: s \rightarrow t+\Dphi,\,  t \rightarrow s-\Dphi \, \, ; \, \, 
S-U \, {\rm exchange}: s \rightarrow -s -t+\Dphi, \, t \rightarrow t \,
\ee 
and compositions thereof. It is most natural to consider the combinations $(s', t', u')\equiv (s, t+\Dphi, -s-t+\Dphi)$
which obey $s'+t'+u'=2\Dphi$ by definition. Crossing symmetry then acts by simple permutations on the three variables $(s', t', u')$. 

This enables us to easily characterise the polynomials in the Mellin variables which are invariant under crossing. 
Since the only linear permutation invariant $(s'+t'+u')$ is a constant, the building blocks are the higher invariants. 
Because of the linear relation above, at quadratic level there is only one independent invariant which can be taken to be $(s't'+t'u'+u's')$. And similarly one at cubic order which can be chosen as $s't'u'$. Thus for crossing symmetric polynomials we can trade the two independent variables, $(s,t)$ for monomials built from the above two lowest order invariants. For convenience we work with 
$\big(\Dphi^2-(s't'+t'u'+u's')\big) = \big(t(s+t)+s(s-\Dphi)\big)$.  
Thus the most general contact witten diagram with a maximum spin $L$ exchanged
can be parametrised as 
\be\label{wittcont}
M^{(c)}_L(s,t)=\sum_{\{m,n\}=0}^{m+n=\frac{L}{2}} a_{mn} \left[s(s+t-\D_\phi)(t+\D_\phi)\right]^m \left[t(s+t)+s(s-\D_\phi)\right]^n\,,
\ee
where the $a_{mn}$'s are constants. We note that this expansion corresponds to a derivative expansion for bulk vertices. The cubic invariant has upto six derivatives and the quadratic invariant four derivatives (each factor of $s$ or $t$ counts as two derivatives). Thus for a maximum spin $L$ we consider terms with upto $3L$ derivatives. 

We also immediately see that the number of independent monomials of spin $L$ (necessarily even for identical scalars) is $\frac{L}{2}+1$. Thus for $L=0$ we have the constant term parametrised by $a_{00}$ and for $L=2$, two terms,  
$a_{01}, a_{10}$. The constant term at $L=0$ corresponds, in position space, to the simplest contact $\Phi^4$ vertex in the bulk which gives the D-function $\bar{D}_{\Dphi, \Dphi,\Dphi,\Dphi}(u,v)$. Similarly, it can be verified that the 
vertex $(\nabla \Phi)^2(\nabla \Phi)^2$ which leads to a D-function as given in Eq.(5.12) of \cite{heemsk} is a linear combination of 
terms with $a_{00}$ and $a_{01}$. This is understandable because the cubic invariant corresponds to six derivatives while the quadratic one to four derivatives. The other vertex which leads to a spin-2 exchange is one with six derivatives and of the form $(\nabla \Phi)^2(\nabla_\mu\nabla_\nu \Phi)^2$ and leads to an involved combination of D-functions (see Eq.(5.15) of \cite{heemsk}). In Mellin space, this term is a simple linear combination of $a_{01}, a_{10}$ and $a_{00}$.  More generally, one can see that the counting of independent terms for any given spin matches exactly with that of \cite{heemsk}. 

\section{Simplifications}

In this section we will use the meromorphic form of the Witten exchange diagrams  to show how they simplify some of the calculations of the Polyakov-Mellin bootstrap. In particular, we will see that this gives a very explicit way to make a  partial $s$-wave decomposition of the $t,u$-channel Witten exchange diagrams. This is very important in the implementation of the Polyakov-Mellin bootstrap where we decompose the crossing symmetric amplitude into such partial waves and extract the residues of the spurious poles at $s=\Dphi$ (more generally at $s=(\Dphi+r)$). We will see that this technical problem is completely solved using the results of this section (and appendix \ref{appd}). See eqs. (\ref{qschann}), (\ref{qcoefft2}) for the answer at $s=\Dphi$ (and eqs.(\ref{qsgen}), (\ref{qtchann}) for the more general case)
These results can be viewed as obtaining an analogue of the crossing kernel, now for Witten diagrams rather than conformal partial waves. It will be interesting to explore the parallels with explicit expressions for the crossing kernel obtained recently in \cite{tar-slght, 6jdsd}.  

\subsection{Decomposition of the exchange Witten diagram in the $s$-channel}

We start with the $s$-channel exchange Witten diagram given in Eq.(\ref{meromwitt}). We want to extract the residue at the double and single poles at $s=\Dphi+r$ and expand it in the orthogonal continuous Hahn Polynomials $Q^{2\D_\phi+2r+\ell'}_{\ell',0}(t)$. It is obvious from the form of the $t$-dependence in Eq.(\ref{meromwitt}), which is through the Mack polynomial ${\widehat P^{(s)}}_{\Delta-h, \ell}(s,t)$ (of degree $\ell$), that we will only have contributions to $Q^{2\D_\phi+2r+\ell'}_{\ell',0}(t)$ with $\ell' \leq \ell$. Thus the scalar exchange diagram contributes only to the $\ell'=0$ channel. The spin 2 exchange to the $\ell'=0,2$ channels etc. Note that this is different from the way in which this decomposition happened in \cite{longpap}. There, in the $s$-channel, only 
$\ell'=\ell$ partial waves contributed. The difference can be attributed to the somewhat different prescription we are adopting here for the exchange diagrams. As discussed in the previous section, the two prescriptions differ by contact diagrams and this is reflected in the additional (finite number of) partial waves that are contributing here. 

We will illustrate how to compute the decomposition for the case of the leading double trace spurious double pole at $s=\Dphi$. The general case (of $s=(\Dphi+r)$ as well as of the single pole) will also straightforwardly follow from the considerations below.  
\be
M_{\Delta,\ell}^{(s)}(s =\Dphi,t) =\sum_{\ell'}q^{(2,s)}_{\D, \ell' |\ell} Q^{2\D_\phi +\ell'}_{\ell',0}(t) \, .
\ee
Since we are setting $s$ to a definite value, the main challenge is the expansion of the Mack polynomial.
For this it will be convenient to use the Dolan-Osborn form given in Appendix \ref{B}, whereby
\be
{\widehat P^{(s)}}_{\Delta-h, \ell}(s,t) =\sum_{m,n \geq 0}^{m+n\leq \ell} \mu_{m,n}^{(\ell)}(\frac{\D-\ell}{2}-s)_m (-t)_n \,,
\ee 
where the coefficients $\mu_{m,n}^{(\ell)}$ are independent of $(s,t)$ and given explicitly in Eq.(\ref{mudef}).
The $t$-dependence on the RHS has a nice expansion (using the orthonormality eq.(\ref{Qorth}))
\be\label{tpochexp}
(-t)_n=\sum_{\ell'=0}^n \c_{\ell'}^{(n)}(s) Q_{\ell',0}^{2s+\ell'}(t)\,,
\ee
with
\be\label{schi1}
\c_{\ell'}^{(n)}\!(s)=(-1)^{\ell'}2^{-\ell'}\frac{\G(2s+2\ell')\G(s+n)^2}{\ell'! \G(\ell'+s)^2\G(2s+n)}\,{}\frac{(-n)_{\ell'}}{(2s+n)_{\ell'}}\,.
\ee
Here we have introduced an $s$ dependence in $Q_{\ell',0}^{2s+\ell'}(t)$ (and therefore in the coefficients  $\c_{\ell'}^{(n)}\!(s)$). This will enable us to consider the more general cases.

Putting it all together (and evaluating the above expressions at $s= \Dphi$) we find 
\be\label{qschann}
q^{(2,s)}_{\D, \ell' |\ell} = \sum_{m,n} \mu_{m,n}^{(\ell)}(\frac{\D-\ell}{2}-\Dphi)_m\chi^{(n)}_{\ell'}(\Dphi)
 \frac{\Gamma^2 \left(\frac{\Delta+\ell }{2}+\Delta_\phi -h\right)\Gamma(2\Dphi-h+\ell)}{\Gamma^2(1+\frac{\Delta-\ell}{2}-\Dphi) } \,.
\ee
One can also use the general form (for arbitrary $s$) of the coefficients in Eqs.(\ref{tpochexp}, \ref{schi1}) to expand around $s=\Dphi$ for the spurious single pole contribution and more generally for $s=\Dphi+r$.
Thus we can write the coefficient of  $Q_{\ell',0}^{2s+\ell'}(t)$ as
\begin{eqnarray}\label{qsgen}
q^{(s)}_{\D, \ell' |\ell}(s) &=& \sum_{m,n} \mu_{m,n}^{(\ell)}(\frac{\D-\ell}{2}-s)_m\chi^{(n)}_{\ell'}(s)
 \frac{\Gamma^2 \left(\frac{\Delta+\ell }{2}+\Delta_\phi -h\right)}{(\frac{\Delta-\ell}{2}-s) 
 \Gamma (\Delta-h +1)} \,\\
&\times & _3F_2\left[\frac{\Delta -\ell}{2}-s,1+\frac{\Delta-\ell }{2}-\Delta_\phi ,1+\frac{\Delta-\ell }{2}-\Delta_\phi ;1+\frac{\Delta-\ell }{2}-s,\Delta-h +1;1\right]\, .\nonumber 
\end{eqnarray}

%{\bf AS: It is important to choose $\s=s$ for the subtracted equations to get rid off higher spin contributions in the epsilon expansion; else you will have to resum their contributions making life very hard. See my comments after 3.25 later as well.}

\subsection{Decomposition of the scalar exchange Witten diagram in the $t$-channel}

To decompose the $t$-channel exchange Witten diagram in the $s$-channel partial waves is more challenging. We will see that the explicit meromorphic form of the Witten diagram helps us make technical progress and solve this problem completely. We first warm up in this subsection by considering the simplest case of scalar exchange. 

We begin with the meromorphic expression for the $s$-channel scalar exchange Witten diagram given in Eq.(\ref{meromwitt}) for $\ell=0$ i.e. 
\be
M_{\Delta,\ell=0}^{(s)}(s,t) = \frac{\Gamma^2 \left(a+2\Delta_\phi -h-1\right)}{(\frac{\Delta}{2}-s) \Gamma (\Delta-h +1)} \,  _3F_2\left[\frac{\Delta}{2}-s, a , a ;1+\frac{\Delta}{2}-s,\Delta-h +1;1\right]\,
\ee
where we have used the fact that $ {\widehat P^{(s)}_{\Delta-h, \ell=0}(s,t)} =1$ in our normalisation  and denoted $a=1+\frac{\Delta}{2}-\Delta_\phi$ for simplicity.

We can now easily transform this to the $t$-channel using the prescription in Eq.(\ref{channexch}). We have
\be
M_{\Delta, \ell =0}^{(t)}(s=\Dphi,t) = \frac{\Gamma^2 \left(a+2\Delta_\phi -h-1\right)}{(a-1-t) \Gamma (\Delta-h +1)} \,  _3F_2\left[a-1-t, a , a ; a-t, \Delta-h +1;1\right]\,.
\ee
For finding out, for instance, the coefficient of the spurious double pole at $s=\Delta_{\phi}$ (see Sec. 4.1.2 of \cite{longpap}) we need to expand this in terms of the continuous Hahn Polynomials $Q^{2\D_\phi +\ell'}_{\ell'}(t)$ which form an orthogonal basis.  
\be
M_{\Delta,\ell=0}^{(t)}(s =\Dphi ,t) =\sum_{\ell'}q^{(2,t)}_{\D, \ell' |\ell=0} Q^{2\D_\phi +\ell'}_{\ell'}(t) \, .
\ee
We can then use the orthogonality relation Eq.(\ref{Qorth}) to read off the coefficient $q^{(2,t)}_{\D, \ell' |\ell=0}$  of the $\ell'th$ partial wave\footnote{The superscript $\lq(2)\rq$ refers to the double pole contribution. One can write a similar expression for the single pole piece. See Sec. 4.1 of \cite{longpap}.}. 
\be\label{qcoeff}
q^{(2,t)}_{\D, \ell' |\ell=0}=\kappa_{\ell'}(\Dphi)^{-1}\int_{-i\infty}^{i\infty}  \frac{dt}{2\pi i} \G^2(\Dphi+t) \G^2(-t) Q^{2\Dphi+\ell'}_{\ell',0}(t)M_{\Delta,\ell=0}^{(t)}(s=\Dphi,t) \, .
\ee
Here $\kappa_{\ell}(\Dphi)$ is a normalisation factor given by Eq.(\ref{kapdef}) for $s=\Dphi$. 
This expression is already a simplification over the ones we had in e.g. Eq. (4.9) of \cite{longpap} where one had an additional intergal over the spectral parameter. 

The expression Eq.(\ref{qcoeff}) can be simplified further as follows. For illustration we further specialise to the case of $\ell'=0$ (which will also be the important case in examples like the $\e$-expansion).  We consider the generalisation for general $\ell'$ in the next subsection.  
It will be useful to first make a transformation Eq.(\ref{3f2t}) to re-express 
\be\label{tchann1simp}
M_{\Delta, \ell =0}^{(t)}(s =\Dphi,t) = \frac{\Gamma \left(2\Delta_\phi -h\right)}{(a-1-t) (a+2\Delta_{\phi}-h-1)} \,  _3F_2\left[1, a, -t,; a-t, a+ 2\Delta_\phi-h ;1\right]\,
\ee 
and then use the identity (see Eq.(\ref{3f2int}))
\be\label{3f2int1}
{}_3F_2[1,a, -t; a-t, a+(2\Delta_{\phi}-h);1]=\frac{\G(a-t)}{\G(a)\G(-t)}\int_0^1 dy\ y^{-t-1}(1-y)^{a-1}{}_2F_1[1, a; a+(2\Delta_{\phi}-h);y]\, . 
\ee
As a result the $t$- integral in Eq.(\ref{qcoeff}), with $\ell'=0$, can now be done explicitly since it takes the simple form 
\begin{eqnarray}\label{tintsimp}
&&\frac{1}{2\pi i}\int dt \ \G(-t)\G(a-t-1)\G^2(\Dphi+t) y^{-t}=\nonumber\\&&\frac{\G^2(\Dphi)\G^2(\Dphi+a-1)}{\G(2\Dphi+a-1)}{}_2F_1[\Dphi ,\Dphi ; 2\Dphi+a-1,1-1/y]\,.
\end{eqnarray}
We can further perform a Pfaff transformation and write
\be
{}_2F_1[s,s;2s+a-1,1-1/y] = y^s {}_2F_1[s,s+a-1;2s+a-1,1-y].
\ee

As a result of all this we have 
\begin{eqnarray}\label{qcoeff2}
q^{(2,t)}_{\D, \ell'=0 |\ell=0} &=& \frac{\G(2\Dphi)\G^2(\Dphi+a-1)}{\G^2(\Dphi)\G(a)\G(2\Dphi+a-1)} \frac{\Gamma \left(2\Delta_\phi -h\right)}{(a+2\Delta_{\phi}-h-1)} \\ 
& \times &\int_{0}^{1}  dy 
 y^{\Dphi-1}(1-y)^{a-1}{}_2F_1[1, a; a+(2\Delta_{\phi}-h);y]{}_2F_1[\Dphi,\Dphi+a-1;2\Dphi+a-1;1-y]  \nonumber\, .
\end{eqnarray}
More general expressions which allow one to write down the residues at the single pole as well as for the case of $s=\Dphi+r$ are given in the next subsection and appendix \ref{appd}. 

As we show, in Appendix \ref{appd}, through similar steps we can arrive at an alternate form  
\begin{eqnarray}\label{qcoeffdi1}
q^{(2,t)}_{\D, \ell'=0 |\ell=0} &=&\frac{\G^2(\frac{\D}{2}+\Dphi-h)}{\G(\D-h+1)}  \int_{0}^{1}  dy \,
 (1-y)^{2\Dphi-h-1}y^{\frac{\D}{2}-1}\\
& \times &{}_2F_1[\frac{\D}{2}+\Dphi-h,\frac{\D}{2}+\Dphi-h,\D-h+1;y]{}_2F_1[\Dphi,\Dphi,2\Dphi;1-y]  \nonumber\, .
\end{eqnarray}

Using these forms, we can explicitly do the finite $y$-integral in full generality. 
%More generally, we can write this integral in terms of a double sum by using the series expansions for the ${}_2F_1$.  
%See Appendix\ref{appd} for more details on this. 
Here we would like to point out that the expression 
for $q^{(2,t)}_{\D, \ell'=0 |\ell=0}$ in the special case where $\Dphi=1$ is given in Eq.(\ref{exactck1}).

Another interesting special case is of $h=1/2,\Dphi=1$ (i.e., $d=1$ case where there are only scalars). If we add in the $s$-channel 
contribution to the $s=\Dphi$ double pole we find these contributions to the double pole residue to be
\begin{eqnarray}
\sum_{\Delta} c_\Delta \big(q^{(2,s)}_{\D, \ell'=0 |\ell=0}+ 2q^{(2,t)}_{\D, \ell'=0 |\ell=0}\big) = \sum_{\Delta} C_\Delta \frac{\G(2\D)}{\G^2(\D)}\frac{\sin^2\frac{\pi \D}{2}}{\pi^2}\left(\psi'(\frac{\D+1}{2})-\psi'(\frac{\D}{2})-\frac{1}{(\D-2)(\D+1)}\right)\,.
\end{eqnarray}
Here $C_\Delta$ is the correctly normalized OPE coefficient - see Eq.(C.7) of \cite{longpap} - and the rational part is the $s$-channel contribution. The same expression has recently been derived by D. Mazac and M. Paulos \cite{mazacpaulos}\footnote{We would like to thank Dalimil Mazac for sharing unpublished notes  with us.} in their extremal functional approach \cite{Mazac1, Mazac:2018mdx}. It will be interesting to explore further the relation between the present approach and theirs. 

In general dimensions, the $q^{(2,s)}_{\Delta,\ell},q^{(2,t)}_{\D, \ell'=0 |\ell}$ behave like $1/\Delta^2$ in the large $\Delta$ limit. If we put in the MFT behaviour for the OPE coefficients then these exchange contributions to the $s=\Dphi$ double pole behaves like $1/\Delta^{2h-2+2\ell}$ in the large $\Delta$ limit. This would imply that for $\ell=0$,  the sum over $\D$ would be divergent, in $d\leq 3$, for theories whose OPE coefficients die off as the MFT. However for $\ell\geq 2$, the sum over $\Delta$ would be convergent with this behaviour for the OPE coefficients. %({\bf RG: Have to put in new dsd reference here?})

%We specialised in the above to $\ell'=0$ to obtain explicit forms of the coefficient of the double pole. More generally, we need to carry out the integral in Eq.(\ref{qcoeff}). We can carry out the steps following Eq.(\ref{qcoeff}) in a similar manner. We use the expression for $Q^{2\Dphi+\ell'}_{\ell',0}(t)$ in the 
%form given in Eq.(\ref{Qdefn2}) where the $t$-dependence can be expressed as a finite sum of terms involving the Pochammers $(-t)_p$. This is because the ${}_3F_2$ function there has only $\ell'$ terms. An additional insertion of $(-t)_p =\frac{\G(p-t)}{\G(-t)}$ in the $t$-integral of Eq.(\ref{tintsimp}) just changes the $\G(-t)$ factor there to 
%$\G(p-t)$. This can also be carried out. The final answer for $q^{(2,t)}_{\D, \ell' |\ell=0}$ can thus be written as a finite sum of terms of the form on the RHS of Eq.(\ref{qcoeff2}). 

\subsection{More general cases}

We specialised in the above to $\ell'=0$ to obtain relatively simple forms of the coefficient of the double pole. The surprise is that the general case can be treated in an explicit way as well leading to more complicated expressions in terms of a ${}_7F_6$ hypergeometric function which has potentially close relations to the $6j$ symbols of the conformal group. 

First we show how to carry out the integral 
%in Eq.(\ref{qcoeff}). We can carry out the steps following Eq.(\ref{qcoeff}) in a similar manner. We use the expression for $Q^{2\Dphi+\ell'}_{\ell',0}(t)$ in the 
%form given in Eq.(\ref{Qdefn2}) where the $t$-dependence can be expressed as a finite sum of terms involving the Pochammers $(-t)_p$. This is because the ${}_3F_2$ function there has only $\ell'$ terms. An additional insertion of $(-t)_p =\frac{\G(p-t)}{\G(-t)}$ in the $t$-integral of Eq.(\ref{tintsimp}) just changes the $\G(-t)$ factor there to 
%$\G(p-t)$. This can also be carried out. The final answer for $q^{(2,t)}_{\D, \ell' |\ell=0}$ can thus be written as a finite sum of terms of the form on the RHS of Eq.(\ref{qcoeff2}). In fact, we can also generalise to the case of 
for $q^{(2,t)}_{\D, \ell' |\ell}$ given by 
\be\label{qcoeff3}
q^{(2,t)}_{\D, \ell' |\ell}=\kappa_{\ell'}(\Dphi)^{-1}\int_{-i\infty}^{i\infty}  \frac{dt}{2\pi i} \G^2(\Dphi+t) \G^2(-t) Q^{2\Dphi+\ell'}_{\ell',0}(t)M_{\Delta,\ell}^{(t)}(s=\Dphi,t) \, .
\ee
Here the $t$-channel block is given by the replacements $s \rightarrow t+\Dphi,\,  t \rightarrow s-\Dphi$ in the $s$-channel expression given in Eq.(\ref{meromwitt})
\begin{eqnarray}
M_{\Delta,\ell}^{(t)}(s=\Dphi,t) &=& {\widehat P^{(s)}_{\Delta-h, \ell}(t+\Dphi, s-\Dphi)} \frac{\Gamma^2 \left(a_\ell+\ell + 2\Delta_\phi -h -1\right)}{(a_\ell-t-1) \Gamma (\Delta-h +1)} \,  \\ \nonumber
&\times & _3F_2\left[a_\ell-1-t, a_\ell , a_\ell ; a_\ell-t, \Delta-h +1;1\right]\, \\ \nonumber
&=& {\widehat P^{(s)}_{\Delta-h, \ell}(t+\Dphi, s-\Dphi)} \frac{\Gamma \left(2\Delta_\phi +\ell -h\right)}{(a_\ell-1-t) (a_\ell +\ell +2\Delta_{\phi}-h-1)} \, \\
&\times&  _3F_2\left[1, a_\ell, -t,; a_\ell -t, a_\ell+ \ell+2\Delta_\phi-h ;1\right] \nonumber
\end{eqnarray}
where $a_\ell =1+ \frac{\D-\ell}{2}-\Dphi$ and we have made the transformation Eq.(\ref{3f2t}) on the ${}_3F_2$ in going to the second equality. This is the generalisation of the form given for $\ell=0$ in Eq.(\ref{tchann1simp}). 

We then follow steps similar to that below Eq.(\ref{tchann1simp}). 
We write the ${}_3F_2$ in integral form 
\be
_3F_2\left[1, a_\ell, -t,; a_\ell -t, a_\ell+ \ell+2\Delta_\phi-h ;1\right] =\frac{\G(a_\ell-t)}{\G(a_\ell)\G(-t)}\int_0^1 dy\ y^{-t-1}(1-y)^{a_\ell-1}{}_2F_1[1, a_\ell; a_\ell+\ell+ (2\Delta_{\phi}-h);y] \, .
\ee
We need to use the expansion for the Mack polynomial in the $t$-channel
\be\label{mackt}
P^{(s)}_{\Delta-h, \ell}(t+\Dphi, s-\Dphi)=\sum_{m,n \geq 0}^{m+n\leq \ell} \mu_{m,n}^{(\ell)}(a_\ell-1-t)_m (\Dphi-s)_n.
\ee
Note that only the $n=0$ term contributes when we evaluate the contribution to the double pole at $s=\Dphi$.
We also use the expression in Eq.(\ref{Qdefn})
\begin{eqnarray}\label{Qtexp}
Q^{2\D_\phi +\ell'}_{\ell'}(t) &= &(-1)^{\ell'}\frac{2^{\ell'} ((\Dphi)_{\ell'})^2}{(2\Dphi+\ell'-1)_{\ell'}}\ {}_3F_2\bigg[\begin{matrix} -\ell',2\Dphi+\ell'-1,-t\\
\Dphi, \Dphi 
\end{matrix};1\bigg] \\ 
&=& (-1)^{\ell'}\frac{2^{\ell'} ((\Dphi)_{\ell'})^2}{(2\Dphi+\ell'-1)_{\ell'}} \sum_{p=0}^{\ell'} \frac{(-\ell')_p(2\Dphi+\ell'-1)_p (-t)_p}{p! ((\Dphi)_p)^2} \,. 
\end{eqnarray}

Bringing together all the $t$-dependence in Eqs.(\ref{mackt}, \ref{Qtexp}), the $t$-integral we need to carry out then  becomes
\begin{eqnarray}\label{tintsimp2}
&&\frac{1}{2\pi i}\int dt \ \G(p-t)\G(a_\ell +m-t-1)\G^2(\Dphi+t) y^{-t}=\nonumber\\&&\frac{\G^2(\Dphi +p)\G^2(\Dphi+m+a_\ell-1)}{\G(2\Dphi+p+m+a_\ell-1)}y^{\Dphi}{}_2F_1[\Dphi +p,\Dphi+m+a_\ell-1 ; 2\Dphi+p+m+a_\ell-1,1-y]\,.\nonumber\\
\end{eqnarray}

Putting everything together we have a pleasing generalisation of Eq.(\ref{qcoeff2})
\begin{eqnarray}\label{qcoefft2}
q^{(2,t)}_{\D, \ell' |\ell}  &=&\frac{2^{-\ell'}}{\ell'!} \frac{\G(2\Dphi+2\ell')}{\G^2(\Dphi+\ell')\G(a_\ell)} \frac{\Gamma \left(2\Delta_\phi +\ell -h\right)}{(a_\ell+\ell+2\Delta_{\phi}-h-1)} \\ \nonumber
&\times & \sum_{p=0}^{\ell'}\sum_{m =0}^{\ell} \mu_{m,0}^{(\ell)} \frac{\G^2(\Dphi+m+a_\ell-1)}{\G(2\Dphi+p+m+a_\ell-1)}\frac{(-\ell')_p(2\Dphi+\ell'-1)_p}{p!} \int_{0}^{1}  dy 
 y^{\Dphi-1}(1-y)^{a_\ell-1} \\ 
& \times & {}_2F_1[1, a_\ell; a_\ell+\ell+(2\Delta_{\phi}-h);y]{}_2F_1[\Dphi+p,\Dphi+m+a_\ell-1;2\Dphi+p+m+a_\ell-1,1-y]  \nonumber\, .
\end{eqnarray}

As we show in Appendix \ref{appd} we can write this in equivalent forms and using that the sum over $p$ can actually be carried out. The $y$-integrand can again be written in terms of a product of  ${}_2F_1$ and ${}_3F_2$ hypergeometric functions - see Eq.(\ref{tchangen}). This further reduces to a product of two  ${}_2F_1$'s. As we will record below in more generality, this $y$-integral can also be carried out. 

In general, to write down the consistency conditions for $s=\Dphi+r$, as well as the simple pole condition, we will need to have a more general expansion 
\be
M_{\Delta,\ell}^{(t)}(s,t) =\sum_{\ell'}c_{\D,\ell}q^{(t)}_{\D, \ell' |\ell}(s) Q^{2s +\ell'}_{\ell'}(t) \, .
\ee
In this case we find, upon using the expression for the Mack polynomial in eq.(\ref{mackt}), the 
generalisation of eq.(\ref{qcoefft2})  
\begin{eqnarray}\label{qcoefft2s}
q^{(t)}_{\D, \ell' |\ell}(s)  &=& \frac{2^{-\ell'}}{\ell'!}\frac{\G(2s+2\ell')}{\G^2(s+\ell')\G(a_\ell)} \frac{\Gamma \left(2\Delta_\phi +\ell -h\right)}{(a_\ell+\ell+2\Delta_{\phi}-h-1)} \\ \nonumber
&\times & \sum_{p=0}^{\ell'}\sum_{n=0}^{\ell}\sum_{m =0}^{\ell-n} \mu_{m,n}^{(\ell)}(\Dphi-s)_n \frac{\G^2(s+m+a_\ell-1)}{\G(2s+p+m+a_\ell-1)}\frac{(-\ell')_p(2s+\ell'-1)_p}{p!} \int_{0}^{1}  dy 
 y^{s-1}(1-y)^{a_\ell-1} \\ 
& \times & {}_2F_1[1, a_\ell; a_\ell+\ell+(2\Delta_{\phi}-h);y]{}_2F_1[s+p,s+m+a_\ell-1;2s+p+m+a_\ell-1,1-y]  \nonumber\, .
\end{eqnarray}

Remarkably, in this most general case too, we can do the $p-$sum and the $y$-integral. 
We simply record the final answer with the details given in Appendix D.
\be\label{qtchann}
q^{(t)}_{\D,\ell'|\ell}(s)=\sum_{m,n}^{\ell} (-1)^{\ell'+m}2^{-\ell'}\mu_{m,n}^{(\ell)} (\Dphi-s)_n (a_\ell)_m^2 \G(2s+2\ell')\G^2(d)\G(\frac{a}{2})\G(a+1)\G^2(1+a-f-b)\tilde W(a;b,c,d,e,f)\,,
\ee
where the parameters $a,b,c$ etc. are given in eq.(\ref{params}) and the $\tilde W$ is the regularized version of a special (``very well poised") ${}_7F_6$ hypergeometric function as defined in eq.(\ref{Wdef}).

As promised, this gives a complete solution to the question of extracting the contribution of the spurious poles in the crossed channels. This answer also reduces (nontrivially) to the various special cases given in Appendix D.1. 

%Notice that the only $s$-dependence comes through the Mack polynomial as expected.

\subsection{Consistency Conditions}

We are now in a position to put together all the pieces and rephrase the consistency conditions 
for the absence of the spurious double trace operators.
 
These consistency conditions then read  
%[[can choose $\s=\Dphi$, however in the epsilon expansion in the subtracted equation we will then get an infinite number of operators contributing at next to leading order itself rather than cancelling off which happens for the choice $\s=s$ so I suggest we use $\s=s$. This appears to be the same thing as choosing $\s=\Dphi$ to obtain one set of equations which we use for $s=\Dphi$ and choosing $\s=\Dphi+r$ for another set of equations from which we choose the $s=\Dphi+r$. There was an explanation for the cancellation given in  the simplev notes]],[[AS: we need to recast the equations below in our notation, I have left this for later to decide if the notation before is still what we continue to use]]
\begin{eqnarray}\label{qcond}
&&\Big(\sum_{\D,\ell}c_{\D, \ell}\big(q_{\D,\ell' | \ell}^{(s)}(s)+2 q_{\D,\ell' |\ell}^{(t)}(s)\big)+\sum_{n,m} a_{n,m}q^{(c)}_{n,m}(s)\Big) |_{s=\D_\phi+r}=0\,,\\
&&\Big( \sum_{\D, \ell}c_{\D, \ell}\partial_s\big(q_{\D,\ell' | \ell}^{(s)}(s)+2 q_{\D, \ell' | \ell}^{(t)}(s)\big) +2 \partial_s q_{\ell'}^{diss}(s)+\sum_{n,m} a_{n,m}\partial_s q^{(c)}_{n,m}(s)\Big)|_{s=\D_\phi+r}=0\,, \nonumber
\end{eqnarray}
where
\be
q_{\ell'}^{diss}(s)=-\frac{(s-\D_\phi)}{\kappa_{\ell'}(s)\G(\D_\phi-s+1)^2} Q_{\ell',0}^{2s+\ell'}(-\D_\phi)\,,
\ee
is the disconnected part arising from the identity exchange as worked out in \cite{longpap}. Also the contact terms $q^{(c)}_{n,m}(s)$ come from the decomposition of the monomials in (\ref{wittcont}) in terms of the orthogonal continuous Hahn polynomials. Note that since the dependence of each monomial in (\ref{wittcont}) is polynomial in $t$, this is, in principle, straightforward. Exact expressions are not very illuminating and can be worked out on a case by case basis.

\section{The Holographic Bootstrap}

In this section we show how the results of Heemskerk et.al. \cite{heemsk} can be reproduced in the Polyakov-Mellin approach to the bootstrap. This will also show the necessity of adding contact Witten diagrams for consistency. One of the interesting features that this example also highlights is the difference between the Polyakov-Mellin bootstrap approach and the conventional AdS/CFT method for constructing four point functions through Witten diagrams in AdS. Despite many superficial similarities we will see that the underlying organisation is subtly different. 
%\subsection{Generalized free fields}

The idea of \cite{heemsk} was to consider a generalised free field theory (GFF) with the following (somewhat artificial) restrictions:
\begin{itemize}
\item
The single trace operator spectrum consists of a single scalar operator $\phi$ whose interactions have a $Z_2$ symmetry. 
\item
There is a maximum spin $L$  to the (double trace) operators which appear in the OPE of two of these scalar operators (to leading order in the GFF coupling $g \ll 1$). 
\end{itemize}

What was then demonstrated was that the solutions to the CFT crossing equations for the four point function of this scalar are in one to one correspondence to a local bulk AdS theory of a scalar field $\Phi$ with only quartic contact interactions with a maximum number of derivatives (which translates into a maximum spin). More precisely, there is a finite dimensional (for fixed $L$) family of perturbative CFTs which can be parametrised (in the bulk AdS) by the different coefficients of the quartic interactions with maximum spin $L$. For this finite dimensional family, one could compute an infinite number of anomalous dimensions $\g_{r,\ell}$ (and corrections to OPE coefficients from the GFF limit) to leading order in $g$. These are for the double trace operators of the  schematic form 
${\cal O}_{r,\ell} \sim \phi \partial^{2r}\partial^{\ell}\phi $
whose dimensions are parametrised as $\Delta_{r,\ell}=2\D_\phi+\ell+2r+\g_{r,\ell}$.
The solutions to the crossing equation in the CFT are parametrised in the same finite dimensional way and the anomalous dimensions agree with that obtained from the bulk calculation. 

We will be making a direct comparison of these results to our framework which is also phrased in terms of AdS building blocks. It is therefore appropriate to say a little bit more about the bulk calculation of \cite{heemsk}. Because of the $Z_2$ symmetry there are no cubic vertices involving the bulk scalar 
$\Phi$. Since in the conventional AdS/CFT dictionary one only considers Witten diagrams of fields corresponding to single trace operators, there are therefore no exchange Witten diagrams to be considered. Recall the scalar $\phi$ is the only single trace operator. However, there are contact quartic diagrams. These give rise to logarithmic dependence on the cross ratio which is interpreted as a signature of the leading contribution of the anomalous dimensions $\g_{r,\ell}$ of the double trace operators ${\cal O}_{r,\ell}$  which appear in a conformal block decomposition of the contact diagram. Thus the whole set of $\g_{r,\ell}$ (all $r \geq 0$ and $\ell \leq L$) as well as the OPE coefficients are linear combinations of the finite number of coefficients of the different quartic interactions (to leading order in $g$). This is therefore a nontrivial check of the holographic bootstrap idea. 

\subsection{Reproducing the Holographic Bootstrap}

%For the generalized free field case we have $\Delta=2\D_\phi+\ell+2r+\g_{r,\ell}$ for all $\ell, r$ with $\ell$ even. To give the usual OPE coefficients as defined for instance in \cite{usprl}, $c_{\Delta,\ell}$ is proportional to $1/\Gamma^2(\D_\phi-\frac{\Delta-\ell}{2})$. We will find it useful to write $c_{\Delta,\ell}=\tilde c_{\Delta,\ell}/\Gamma^2(\D_\phi-\frac{\Delta-\ell}{2})$ in what follows. Further in what follows $r+q=n$.  

Our aim now is to reproduce the above results using the Polyakov-Mellin bootstrap formalism. 
In other words, we expand the four point function of the GFF as in Eq.(\ref{witten}). Note that we are now allowing contact diagrams as well. The key differences with the usual holographic prescription bear repetition. Firstly, we are summing over {\it all} conformal primaries -- thus including all the double trace operators and so on. In particular, the net amplitude in Mellin space will have physical double trace poles at  $\Delta_{phys}=2\D_\phi+\ell+2r+\g_{r,\ell}$. Therefore, in particular, we will be including exchange diagrams (with physical double trace operators) unlike in the usual holographic bootstrap. 
Secondly, and relatedly, we require the final answer to not have the spurious double trace poles at the values $\Delta_{sp}=2\D_\phi+\ell+2r$ and we impose this as a condition on the full amplitude. As we will show soon, this is achieved by cancelling off the log terms coming from the contact diagrams with the corresponding pieces coming from the exchange diagrams. Note that the spurious double trace poles are $O(g)$ away from the physical double trace poles and we carry out the small $g$ expansion only on the residues and {\it not} on the location of the poles (which would mix these two different sets of poles).    

Despite these differences, it can be checked\footnote{Mathematica notebook can be provided on request.} that one reproduces all the  $\g_{r,\ell}$ and the corrections to the OPE coefficients exactly as in \cite{heemsk} if one adds the same contact diagrams. Remarkably, the coefficients of these contact diagrams are exactly what one finds in the holographic bootstrap as parametrising the solutions of the crossing equation. 

One can however make a general argument for the agreement of the two approaches as follows. In the usual holographic bootstrap, the four point amplitude in the bulk is given by the manifestly crossing symmetric contact diagrams (which contribute to a given spin) with arbitrary coefficients (which are proportional to $g$ to leading order). From the point of view of the field theory this is reproduced by the usual conformal block expansion, in say the $s$-channel
\be\label{schancbl}
{\cal A}(u,v) = \sum_{\Delta,\ell} c_{\Delta,\ell} \int \dsdt\, u^s v^t B^{(s)}_{\Delta,\ell}(s,t) \rho_{\Dphi}(s,t) \, .
\ee
Here the OPE coefficients $c_{\Delta,\ell}$ and the spectrum $\D$ of physical primaries are fixed by the crossing symmetry equation -- which determines these in terms of the same arbitrary coefficients as in the bulk. Note that $B^{(s)}_{\D,\ell}(s,t)$ is the Mellin space form for the usual conformal block $G^{(s)}_{\D,\ell}(u,v)$. To leading order in $g$, the contribution in the sum comes only from the double trace operators $\D_{phys}$ -- the single trace operator does not contribute since it is prohibited by the $Z_2$ symmetry. The result of the holographic bootstrap is that the above contribution from the double trace operators agrees to leading order in $g$ with the bulk contact diagram terms. 

The crux of the present argument is that we can relate the $s$-channel conformal block expansion in eq.(\ref{schancbl}) to the $s$-channel Witten diagram expansion for the same double trace operators. 
%The way we are defining the Witten diagram basis (for the s-channel part) is simply 
%\be
%\sum_{\Delta,\ell} c_{\Delta,\ell} \int ds dt\, u^s v^t W^{(s)}_{\Delta,\ell}(s,t) \mu(s,t)\,,
%\ee
%{\it i.e.,} the same integrand as eq.(\ref{usual}) without the $\sin^2$ factors. Now we will have spurious poles at $s=\D_\phi+n$ as we know. 
This is because, as discussed in the introduction, we can replace in eq.(\ref{schancbl})
\be
B^{(s)}_{\Delta,\ell}(s,t) \rightarrow W^{(s)}_{\Delta,\ell}(s,t) \frac{\sin^2 \pi(\Delta_\phi-s)}{\sin^2 \pi(\Delta_\phi-\frac{\Delta}{2})}.
\ee 
%({\bf RG: Is there a choice of witten exchange diagrams implicit in this? AS: We discussed this--yes, we are only focusing on the meromorphic piece. BTW, in the equation below we can also replace $W^{(s)}\rightarrow W^{(s)}+W^{(t)}+W^{(u)}$ closing the $s$ contour on the right, since the crossed channels will not have any s poles to pick up. })
It can then be easily checked (eg. on mathematica) that
\be
W^{(s)}_{\Delta,\ell}(s,t) \frac{\sin^2 \pi(\Delta_\phi-s)}{\sin^2 \pi(\Delta_\phi-\frac{\Delta}{2})}\rho_{\Dphi}(s,t)
\ee
at $s=\frac{\Delta-\ell}{2}+q=\Delta_\phi+n+\gamma_{r,\ell}/2$ (which is the usual contribution to the amplitude from the physical double trace operators)\footnote{Note that $n=q+r$.} 
and  $W^{(s)}_{\Delta,\ell}(s,t) \rho_{\Dphi}(s,t)$
at $s=\D_\phi+n$ (which are the spurious double trace poles) have the same residues to linear order in $\gamma_{r,\ell}$ i.e. $\propto g$, upto an overall sign. This establishes that the usual s-channel expansion in eq.(\ref{schancbl}) and the Witten basis s-channel will give the same results for the coefficients of $u^{\D_\phi+n}\log u$ and $u^{\D_\phi+n}$ upto an overall sign. The log pieces in the former were interpreted as corrections from anomalous dimensions and the holographic bootstrap confirmed that these matched with the contact Witten diagram contributions. This near cancellation of the residues of the spurious double poles with the physical double pole contributions is the appearance under another guise of the statement that the Mellin measure $\rho_{\Dphi}(s,t)$ correctly accounts for, in a GFF to leading order in $g$, the contribution of physical double trace operators. 

As we will argue in the next para, the crossed channel exchange Witten diagrams do not contribute to 
leading order in $g$. Thus to cancel the contributions of the spurious poles, in the Polyakov-Mellin approach, from the $s$-channel exchange Witten diagrams we need to add contact terms. Putting this together with the conclusion of the previous para we see that the Witten $s$-channel diagrams when added to the {\it exact same} contact diagram contributions demanded by the holographic bootstrap, will have a a nett vanishing contribution from the spurious poles at $s=\D_\phi+n$.  Thus the Polyakov-Mellin bootstrap conditions are fulfilled precisely when the same contact diagram contributions are added as in the holographic bootstrap. 

%and hence give the same results as in \cite{pol} when the appropriate contact terms are added to the latter case. This provides a straightforward proof of the results found in \cite{dgs} and extends it to all $s=\D_\phi+n$.

What remains to be argued is that the crossed channel contributions are of higher order in $g$. 
%The same argument can be shown to hold in the crossed channel as well. Now there is a difference in the expansions of the crossed channels in the continuous Hahn polynomials that must be pointed out. 
For the $t$-channel Witten diagram, when we do the t-integral to extract the $q^{(t)}(s)$ as in the previous section, then both the physical double trace pole and the spurious poles (now in $t$) contribute and there is a relative cancellation between these at $O(g)$. This follows from the argument presented earlier in the $s$-channel \footnote{Note that the crossed channel expressions are obtained by the replacement $s\rightarrow t+\D_\phi, t\rightarrow s-\D_\phi$.  Following Caron-huot  \cite{caronhuot}, we define the double discontinuity of $v^t$ as ${\rm dDisc~}v^t= v^t\, \sin^2\pi t$.  The above way of writing the usual t-channel block and the Witten basis t-channel block then makes it obvious that the double discontinuity of the two when evaluated around $v=0$ are the same. This is of course the statement that double trace operators do not contribute to the double discontinuity.}. Therefore the Witten basis $q^{(t)}(s)$ begins at $O(g^2)$ as claimed above and does not affect the Polyakov-Mellin bootstrap condition to the leading order. This justifies the conclusion that it is the contact diagrams (with the same bulk coefficients as in the Holographic bootstrap) which must cancel against the $s$-channel exchange to give a consistent and indeed, correct, solution. We therefore see that we are forced to add contact Witten diagrams to consistently impose the Polyakov-Mellin bootstrap conditions.

\section{The Epsilon Expansion}
In our previous papers \cite{usprl, longpap}, we showed how to extract anomalous dimensions and OPE coefficients for double field operators at the Wilson-Fisher fixed point at $d=4-\e$ up to cubic order in $\e$ for spinning operators and to quadratic order for the scalar $\phi^2$ operator. Our present approach illuminates those results afresh and also enables us to extract the anomalous dimension and OPE coefficients of the twist-4 scalar and the twist-4 spin-2 operator, both of which are non-degenerate and for which no bootstrap derivation exists so far. We will outline the steps below for the $\ell=0$ case; the non-zero spin case can be dealt with similarly. 
%(for some useful intermediate algebraic steps, the reader is referred to appendix ..).

The starting point is eq.(\ref{qcoeffdi}). We can proceed in two ways, both of which give the same result. We can use the double sum form in eq.(\ref{tu0}) or expand the ${}_2F_1$'s in eq.(\ref{qcoeffdi}) and carry out the $y$-integral\footnote{Yet another way which leads to the same results is to start with the ${}_7F_6$ form, write it as a difference of two well balanced ${}_4F_3$'s and then epsilon expand these using $Hypexp$.}. Let us outline the second approach as the form of the integrand is suggestive and is closer to the integrals involved in \cite{caronhuot, aldayeps}. First we give the expansions of the hypergeometric functions. These can be obtained using the Mathematica package {\it Hypexp} \cite{hep-ph/0507094}.
Writing $\Dphi=1+\delta_\phi$ we find
\begin{eqnarray}
{}_2F_1[\Dphi,\Dphi,2\Dphi,1-y]&=& \frac{\log y}{y-1}+\frac{2\delta_\phi}{y-1}\left(Li_2(1-y)+\log y\right)\nonumber \\
&+&\frac{\delta_\phi^2}{3(y-1)}\bigg[(\zeta_2-3\log y \log(1-y))\log y-3(\log y-4)Li_2(1-y)\nonumber\\&&~~~~~~~~~~~~~- 12Li_3(1-y)-6 Li_3(y)+6 \zeta_3\bigg]+O(\delta_\phi^3)\,.
\end{eqnarray}
 Next we write $\frac{\D}{2}+\Dphi-h=\delta, \D-h+1=1+\tilde \delta$ and find
 \be
 {}_2F_1[\frac{\D}{2}+\Dphi-h,\frac{\D}{2}+\Dphi-h,\D-h+1;y]=1+\delta^2 Li_2(y)+O(\delta^3,\tilde\delta^3)\,.
 \ee
 Using these we can next carry out the $y$-integral in eq.(\ref{qcoeffdi}) - note that we do not expand the remaining $y$ factors in the integral. While these integrals are somewhat tedious to carry out, the method is straightforward. We need to be careful to expand the product of the ${}_2F_1$'s in the integral to one higher order than what we are interested in as after carrying out the $y$-integral, there is an enhancement in the order. To illustrate this point, let us work out the leading order contribution. In order to be consistent with the normalizations in our previous papers, our blocks have to be multiplied by the following factor:
 \be
 {\mathcal N}_{\Delta,\ell}=\frac{2^\ell (\D+\ell-1)\G^2(\D+\ell-1)\G(\D-h+1)}{\G(\D-1)\G^4(\frac{\D+\ell}{2})\G^2(\Dphi-\frac{\D-\ell}{2})\G^2(\Dphi-\frac{2h-\D-\ell}{2})}\,.
 \ee
Let us denote the OPE coefficient for $\phi^2$ as $C_0=\sum_i C_0^{(i)}\e^i$ and the scaling dimension of $\phi^2$ to be $\Delta_0=2+\sum \delta_0^{(i)}\e^i$. We further write the scaling dimension of the external scalar as $\D_\phi=1+\sum \delta_\phi^{(i)}\e^i$ and $h=d/2=2-\e/2$. The $y$-integral after expanding the hypergeometric functions reads:
\begin{eqnarray}
&&-\e^2 (\delta_\phi^{(1)}-\frac{\delta_0^{(1)}}{2})^2\int_{0}^1 dy \, (1-y)^{2\Dphi-h-2} y^{\frac{\D}{2}-1} \log(y)\nonumber\\&=&\e^2 (\delta_\phi^{(1)}-\frac{\delta_0^{(1)}}{2})^2\frac{\G(2\Dphi-h-1)\G(\frac{\D}{2})}{\G(2\Dphi+\frac{\D}{2}-h-1)}\left(\psi(2\Dphi-h+\frac{\D}{2}-1)-\psi(\frac{\D}{2})\right)\nonumber\\
&=& \frac{(\delta_0^{(1)}-2\delta_\phi^{(1)})^2}{2(1+4\delta_\phi^{(1)})}\e+O(\e^2)\,.
\end{eqnarray}
In a similar manner we can carry out the $y$-integrals systematically in the $\e$-expansion. The results are identical to our previous papers. The double summation form in eq.(\ref{tu0}) makes carrying out the epsilon expansion on Mathematica\footnote{A mathematica notebook can be made available on request.} somewhat easier but both approaches give identical results.

\subsection{New results}

Now with the current machinery developed in this paper, we can look at the consistency conditions for any $s=\Dphi+r$. For instance, looking at the conditions systematically using the techniques above we find the following:
\begin{itemize}
\item New operators start contributing at $\e^3$ order for the $\ell=0$, $s=\Dphi$ equation. By making the mild assumption that higher twist  (twist $\geq$ 4) operators have their OPE coefficients beginning at $O(\e^2)$, we find that all higher twist scalar operators can in principle contribute to this order. Higher twist spinning operators contribute from $\e^4$ with the same assumption. At $\e^3$ order we also need a new piece of information, namely  $\delta_0^{(3)}$, which we have not determined yet. 
\item The situation is similar for the $s=\Dphi+1$ equation. Here the $\ell=2$ (stress tensor) begins to contribute from $\e^2$ order. We can put in the information for this operator which we obtain from the $\ell'=2$ equations. However, except for the twist-4 scalar, all higher twist scalars contribute the same as in the $s=\Dphi$ equation. Also, quite nicely, the dependence on $\delta_0^{(3)}$ cancels out. Therefore, taking the difference of these equations will get rid of all higher twist scalars except the $\phi^4$ operator. 
\item By looking at the difference equation, we find that the dimension of $\phi^4$ is $\Delta_{4,0}=4+O(\e^2)$, or in other words the anomalous dimension is $2\e$. Further the single pole condition gives the OPE coefficient squared $c_{\phi\phi\phi^4}$ to be $\e^2/54$ in our normalizations.
\item By considering the same difference equations for $\ell=2$ and using the fact that the spin-2 twist-4 operator is non-degenerate, we find that $\Delta_{6,2}=6-\frac{5}{9}\e$ and its OPE coefficient squared is $\e^2/1440$. 
\item By iterating this and going to $s=\Dphi+r$ we find that for consistency the OPE coefficients for operators with twist $>4$ must begin at $O(\e^4)$. 
\end{itemize}

Now using the last piece of information, we can look at the $s=\Dphi$, $\ell=0$ condition again by putting in the contribution from the $\phi^4$ operator. According to the discussion above, the $\phi^2$ and $\phi^4$ operators are all that will contribute up to $\e^3$ order. We find that the resulting Polyakov-Mellin bootstrap condition applied naively using only the exchange diagram gives
\be
\delta_0^{(3)}=-\frac{4247}{17496}\,,
\ee
instead of the expected Feynman diagram result
\be
\delta_0^{(3)}=\frac{937}{17496}-\frac{4}{27}\zeta_3\, .
\ee
Thus we see that without adding contact diagrams which can, in principle, contribute at this order, we find a discrepancy in $\delta_0^{(3)}$  which is simply $4(2-\zeta_3)/27 $. We see that not adding the contact diagrams leads to a discrepancy with known results. 
 
\subsection{Contact terms in the epsilon expansion}

If the number of contact terms that need to be added are finite, then by taking linear combinations of the equations, one can hope to eliminate the unknown parameters. To illustrate this suppose that the only contact term needed in the context of the epsilon expansion is the $a_{00}$, i.e., the constant term. Then this will only affect the $\ell'=0$ conditions. In particular by subtracting pairs of equations for $\ell'=0$ and $s=\D_\phi+r_1$ and $s=\D_\phi+r_2$, one can eliminate $a_{00}$ from the conditions. 
It turns out, as we will argue below, that this line of reasoning can demonstrate that $a_{00}$ begins at $O(\e^3)$ which explains our earlier findings as well. 
We will find that the simple pole cancellation condition will need the OPE coefficients to leading order to be that in the MFT. So let us focus on the double pole cancellation conditions.

Let us denote the OPE coefficient for $\phi^2$ as $C_0=\sum_i C_0^{(i)}\e^i$ and the scaling dimension of $\phi^2$ to be $\Delta_0=2+\sum \delta_0^{(i)}\e^i$. We further write the scaling dimension of the external scalar as $\D_\phi=1+\sum \delta_\phi^{(i)}\e^i$. We will also need the stress tensor whose scaling dimension is $\D=4-\e$ and OPE coefficient is $C_2=\sum_i C_2^{(i)}\e^i$.

First for the $\ell'=0$, $s=\D_\phi$ condition eq.(\ref{qcond}) to hold at $O(\e)$ we easily find that $\delta_\phi^{(1)}=-1/2$. Imposing eq.(\ref{qcond}) to hold at $O(\e^2)$ we need
\be
\frac{5}{16}\left(C_0^{(0)}(1+(\delta_0^{(1)})^2)-72 C_2^{(0)}\delta_\phi^{(2)}\right)=0\,.
\ee
The difference between the $s=\D_\phi$ and $s=\D_\phi+1$ condition to leading order in $\epsilon$ leads to
\be
\frac{1}{12}\left(-C_0^{(0)}(1+\delta_0^{(1)})(7+10\delta_0^{(1)})+72 C_2^{(0)}\delta_\phi^{(2)}\right)=0\,.
\ee
Using the simple pole cancellation condition results  $C_0^{(0)}=2$ and $C_2^{(0)}=1/3$ we find that 
\be
\delta_0^{(1)}=-1, \delta_\phi^{(2)}=0 \,,\quad {\rm or}\quad \delta_0^{(1)}=-\frac{2}{3}, \delta_\phi^{(2)}=\frac{1}{108}\,.
\ee

This is how we recover the free theory and the Wilson-Fisher results. Notice that by taking differences we do not lose anything and the results are unambiguously fixed to leading order for $\phi^2$ and second order for $\phi$. Now we can go back and plug the solution into the $s=\D_\phi$ equation without contact term to find that it is satisfied at $O(\e)$. This automatically means that the contact term must begin at least at $O(\e^2)$ for consistency. The reason why this worked so well at this order is because of an implicit assumption that only twist-2 operators show up and the contribution from higher twists is suppressed. 
%{\bf AS: here is where I am stuck now. Without arguing that the contact term begins at $O(\e^3)$ we will not be able to recover our previous results at next order since the difference equation becomes sensitive to the $\phi^4$ operator at next order. We can probably argue that since the original condition began at $O(\e)$, the contact term must begin two orders higher..... another way would be to assume (like we assumed that only twist 2 operators contributed with OPE of $O(1)$) that only twist 4 operators would contribute with OPE of $O(\e^2)$ so that the only scalar that contributes is $\phi^4$ at the next order; then using this we can show that the contact term must begin at $O(\e^3)$. So assuming this--} 
The contact term contribution to individual $s=\D_\phi+r$ equation hence begins at $O(\e^3)$. 

To explicitly see what the problem is, in going to higher orders, consider the double pole cancellation conditions that arise from $s=\D_\phi+r$ to quadratic order in $\e^2$. Including the effect of a constant contact interaction arising at $\e^2$ and parametrizing the unknown coefficient by $\chi$ we get 
$$
\frac{19}{162}-\delta_0^{(2)}+\chi=0\,.
$$
From the $\ell'=2$  condition we get
\be
11+216\delta_0^{(2)}-3888\delta_\phi^{(3)}=0\,,
\ee
and finally from the difference between $s=\D_\phi+1$ and $s=\D_\phi$ we get
\be
103+108\delta_{4,0}^{(1)}-1188\delta_0^{(2)}+3888\delta_\phi^{(3)}=0\,.
\ee
Here we have plugged in the leading and next to leading order OPE coefficient results. These follow from the simple pole cancellation condition that only needed the solution at previous order. Further the $\phi^4$ operator starts showing up in the difference equation and its scaling dimension is parametrized by $\Delta_{\phi^4}=4+\sum_n \delta_{4,0}^{(n)}\e^n$ and its leading OPE coefficient is given by $\e^2/54$. Thus we land up having $(\delta_0^{(2)},\chi,\delta_\phi^{(3)},\delta_{4,0}^{(1)})$ as unknowns but only 3 equations. If we assume $\chi=0$ then we recover exactly the solutions expected namely
\be
\delta_0^{(2)}=\frac{19}{162},\quad \delta_\phi^{(3)}=\frac{109}{11664}\,,\quad \delta_{4,0}^{(1)}=0\, .
\ee

In other words, the fact that we reproduce the correct answers with $\chi=0$ indicates that our assumption that only twist 4 operators would contribute to the OPE coefficients at $O(\e^2)$ so that the only scalar that contributes is $\phi^4$ at the next order, is a consistent one. It would be desirable to have an independent argument which would ensure that consistency requires this. 
%but then it would appear that we need an independent argument for this which does not follow from the self-consistency of the equations unlike what happened in the previous order. 
Note that considering other equations will not help as we will need to introduce new unknown parameters for new operators coming in. 
%Note that this problem of more unknowns than conditions will plague the resolution proposed by Mazac and Paulos as well and hence their resolution cannot be correct for the $\epsilon$ expansion. In fact in their analysis the will miss the $s=\D_\phi+r$ conditions altogether as according to them we can only work with difference equations. [[{\bf AS: Dalimil is aware of this potential problem as he was saying that they do not know what to do yet when new higher twist operators show up in the analysis. A possible way out is to consider other correlators as well in their approach. }]]

\section{Discussion}

In this paper we have elaborated on the crossing symmetric formalism introduced by Polyakov and recast in Mellin space in \cite{usprl, longpap} (and further applied in several contexts \cite{dks, dgs, golden, dk}). In the process, we have enormously simplified the key ingredients in the approach. By using the meromorphic part of the crossing symmetric exchange Witten diagrams, we worked out explicitly the expressions when each channel is expanded in terms of continuous Hahn polynomials. We found that the crossed channel expression can be simplified enormously and the final form is a finite sum of very well poised ${}_7F_6$ hypergeometric functions which can also be written as a difference of well balanced ${}_4F_3$'s, which admit various analytic continuations. The connections that arise to the usual crossing kernel and $6j$ symbols for the conformal group are worth exploring\footnote{An additional subtlety must be kept in mind when making comparisons to the usual crossing kernels. We have found the contribution to the partial wave $Q^{2s+\ell'}_{\ell', 0}(t)$ of the (direct and) crossed channels. Evaluated at the spurious double trace poles $s=\Dphi+r$, this gives not only the contribution of the spurious double trace primaries with dimension $2\Dphi+2r+\ell'$ but also from the descendants of other spurious double trace primaries with dimension $2\Dphi+2r'+\ell'$ (with $r'< r$). Disentangling the latter contributions from the former is a finite well defined exercise which we do not explicitly carry out here.} 

%\footnote{In the current paper, after expanding the s-channel in terms of the continuous Hahn polynomials, we obtain a sum over primary operators as in eq. (3.6). Due to the pole at $s=(\D-\ell)/2+q$, there is an enhancement for $\D=2\Dphi+\ell+2n-2q+\g_{n,\ell}$, $q\leq n$, for the $s=\Dphi+n$ consistency condition at leading order in $\g_{n,\ell}$. Typically these operators will all contribute and one will have to disentangle their contribution--for instance, by considering linear combination of the consistency conditions as was done to extract the anomalous dimension of $\phi^4$ in the epsilon expansion. }.

We further parametrized the potential contact terms in the basis in a suitable way and demonstrated their need by studying the holographic bootstrap as well as the epsilon expansion. The most pressing outstanding question is to come up with a physical principle to constrain these contact terms. It would have been desirable if the self-consistency of the constraint equations themselves dictated what form the contact terms need to take. For instance, in the holographic bootstrap we saw that without adding contact terms it was not possible to deform the theory away from the mean field theory leading order result. In the epsilon expansion example, we were less fortunate. We concluded that there must be contact terms by comparing with the known Feynman diagram answer for the anomalous dimension of $\phi^2$. {\it A priori} there did not seem to be anything inconsistent with the consistency conditions at the order at which we were working. For $d=1$ there appeared to be a problem with the convergence of the $\ell'=0$ constraint and hence appeared to necessitate the introduction of contact terms which can be found \cite{mazacpaulos} using the technique developed in \cite{Mazac:2018mdx}. Effectively this means, in our present framework, working with pairwise differences of the original $\ell'=0, s=\Dphi+r$ constraints.  While working with the difference equations removed the discrepancy we found with the Feynman diagram result, this approach introduced an ambiguity which needed further input to fix.  Moreover, in the epsilon expansion we did not encounter a convergence problem at least at the order we were working; hence, {\it a priori} there was no good reason to work with just the difference equations. Thus it seems that the physical principle dictating the existence of contact terms in the Witten diagram basis is more than just convergence.  It is possible that some extra assumption about how the full Mellin amplitude behaves at large $s,t$ is also needed and the Feynman diagram approach implicitly uses this.

Another idea that may be fruitful to investigate is the momentum space dispersion relation approach advocated in Polyakov's original paper \cite{polya} and investigated in \cite{KSAS}. Using that approach, a bit mysteriously, one lands up with the correct rational part of the anomalous dimension at $O(\e^3)$ \cite{unpublished} by retaining only the scalar $\phi^2$ in the crossed-channel.  Thus asking how the Mellin space approach and the momentum space approach are consistent with one another may fix the contact term ambiguities. Recent progress in this direction include \cite{Isono:2018rrb} but the issue has not been resolved. A direct question then is: How does one implement the momentum space dispersion relation, which arises out of unitarity constraints, in Mellin space?

Finally, a question of interest clearly is: How is the Polyakov-Mellin bootstrap related to the usual approach? Recently, in \cite{aldayeps}, Alday and collaborators have obtained the $\e^4$ answer for anomalous dimensions of higher spin double field operators using Caron-Huot's inversion formula and by assuming that the double discontinuity of the full correlator can be expanded in terms of a basis that respects pure transcendentality (currently there is no bootstrap derivation for this assumption beyond arguing that leading orders seem to obey it). Furthermore, by extending the answer to $\ell=0$, they can also obtain the $O(\e^3)$ anomalous dimension for $\phi^2$. However, the calculational details in that approach seem to be quite different compared to what the Polyakov-Mellin bootstrap uses\footnote{We thank Fernando Alday for discussions on this.}. For instance, to determine the $O(\e^2)$ anomalous dimension for $\phi^2$, the Polyakov-Mellin bootstrap needs to use only the existence of a conserved stress tensor and a $Z_2$ symmetry. Using this, the dimension of $\phi$ is fixed up to $O(\e^3)$ and the dimension of $\phi^2$ up to $O(\e^2)$. On the other hand, the inversion formula seems to need information about the anomalous dimension of higher spin double field operators up to $O(\e^4)$  as well as twist 4 operators for the same purpose. Thus, the consistency conditions seem to have packaged information differently in the two approaches. It will be highly desirable to understand the parallels and differences between the usual approach and the Polyakov-Mellin approach and provide a direct derivation of the pure transcendentality assumption in \cite{aldayeps}.

\section*{Acknowledgments}
We thank Fernando Alday, Parijat Dey, Kausik Ghosh, Apratim Kaviraj, Hugh Osborn, Eric Perlmutter, Leonardo Rastelli, David Simmons-Duffin, Ahmadullah Zahed and especially Dalimil Mazac and Miguel Paulos for numerous useful discussions. 
R. G.'s research was partially suppported by the J. C. Bose Fellowship of the SERB and more broadly by the Indian public's generous funding of basic sciences. 
A.S. acknowledges support from a DST Swarnajayanti Fellowship Award DST/SJF/PSA-01/2013-14. 
R. G. would like to acknowledge hospitality of the Simons bootstrap workshop at Caltech where a preliminary version of these results were presented. A.S. acknowledges hospitality of the Simons bootstrap workshop at Azores where some of these results were presented.
We would both like to also  thank the organisers of the Okinawa Conformal Bootstrap workshop at OIST for providing a stimulating setting for many discussions. 

\appendix
\section{Continuous Hahn Polynomials}\label{A}
The continuous Hahn  polynomials $Q_{\ell,0}^{2s+\ell}(t)$ are defined as
\be\label{Qdefn}
Q^{2s+\ell}_{\ell,0}(t)=\frac{2^\ell ((s)_\ell)^2}{(2s+\ell-1)_\ell}\ {}_3F_2\bigg[\begin{matrix} -\ell,2s+\ell-1,s+t\\
s, s
\end{matrix};1\bigg]\,.
\ee
Under $t\rightarrow -s-t$, this is symmetric up to an overall $(-1)^\ell$.
Namely,
\be\label{Qdefn2}
Q^{2s+\ell}_{\ell,0}(t)=(-1)^{\ell}\frac{2^\ell ((s)_\ell)^2}{(2s+\ell-1)_\ell}\ {}_3F_2\bigg[\begin{matrix} -\ell,2s+\ell-1,-t\\
s, s
\end{matrix};1\bigg]\,.
\ee
The orthonormality condition for these  $Q_{\ell,0}$ polynomials is given by \cite{AAR}
\be\label{Qorth}
\frac{1}{2\pi i}\int_{-i\infty}^{i\infty}\G^2(s+t) \G^2(-t) Q^{2s+\ell}_{\ell,0}(t)Q^{2s+\ell '}_{\ell',0}(t)=\kappa_{\ell}(s)\d_{\ell,\ell'}\,,
\ee
where,
\be\label{kapdef}
\kappa_{\ell}(s)=\frac{4^\ell (-1)^\ell \ell! \G^4(\ell+s)\G(2s+\ell-1)}{\G(2s+2\ell)\G(2s+2\ell-1)}\,.
\ee

\section{Mack polynomials: conventions and properties}\label{B}

The explicit expression for the Mack polynomials can be found in e.g.\cite{mack,dolanosborn2,regge}.
A convenient form for these polynomials was given in eq. (3.47) in \cite{dolanosborn2}. Our normalizations are:
%\begin{align}\label{sumt2}
%\begin{split}
%P^{(s)}_{\nu,\ell}(s,t)&=\frac{2^{-\ell}}{(h-1)_\ell}\sum_{m+n+p+q=\ell}\frac{\ell!}{m!n!p!q!}(-1)^{p+n}(2\bar\l_2+\ell-1)_{\ell-q}(2\l_2+\ell-1)_{n}(\bar\l_1-q)_q^2(\l_1-m)_m^2 \\
%&\times (2h-2+\ell+n-q)_q (h-1)_{\ell-q}(h-1+n)_p (\l_2-s)_{p+q}(-t)_n\,.
%\end{split}
%\end{align}
%This form is convenient for several reasons. First, the $s,t$ dependence is considerably simpler than the other form given above. Second, the form while hiding symmetries such as $\nu\rightarrow -\nu$ makes a simplification for $s=\lambda_2$ quite manifest. In this case, the restricted sum sets $p=q=0$ and reduces to a single sum which can be checked to be of a ${}_3F_2$ form quite easily. 
%To shorten notation we write
\be
 \widehat P^{(s)}_{\D-h,\ell}(s,t) =\sum_{m,n} \mu_{m,n}^{(\ell)}(\frac{\D-\ell}{2}-s)_m (-t)_n=(-1)^\ell\sum_{m,n} \mu_{m,n}^{(\ell)}(\frac{\D-\ell}{2}-s)_m (s+t)_n\,,
 \ee
 where $\mu_{m,n}^{(\ell)}$'s are defined in eq.(\ref{mudef}) below and the second equality comes from using the symmetry under $s\rightarrow s, t\rightarrow -s-t$ of the Mack polynomial. It is the first form we should use when doing the $s,t$ channels and the second form that we should start with when going to the $u$-channel. In the $u$-channel we will need to expand $(s+t)_m$ and so the $t,u$ channels will become same if we set $\sigma=s$ in the basis $Q_{\ell,0}^{2\s+\ell}(t)$ which follows from the invariance (up to an overall $(-1)^\ell$ and since $\ell$ is even for identical scalars this is unity) of the $Q_{\ell,0}^{2s+\ell}(t)$'s under $s\rightarrow s, t\rightarrow -s-t$.  The $\mu_{m,n}^{(\ell)}$'s are defined via
 \begin{eqnarray}\label{mudef}
 \mu_{m,n}^{(\ell)}&=&2^{-\ell} \frac{(-1)^{m+n}\ell!}{m! n! (\ell-m-n)!}(\l_1-m)_m (\l_2+n)_{\ell-n} (\l_2+m+n)_{\ell-m-n}(\ell+h-1)_{-m}(\ell+\D-1)_{n-\ell}  \nonumber \\
 &\times& {}_4F_3[-m,1-h+\l_2,1-h+\l_2,n-1+\D;2-2h+2\l_2,\l_1-m,\l_2+n;1]\,.\nonumber\\
 \end{eqnarray}
 Here $\lambda_1=(\D+\ell)/2, \lambda_2=(\D-\ell)/2$ and $h=d/2$ where $d$ is the number of spacetime dimensions. Further, the last ${}_4F_3$ is a well-balanced one. 
 We make a note of the following $\mathbf{Z}_2$ symmetries for the Mack polynomials:
\begin{itemize}
\item $T_1$: For any $d$, $s\rightarrow s$, $t\rightarrow -s-t$ with $(-1)^\ell$.
\item $T_2$: In addition, for $d=2$, $s\rightarrow -s$, $t\rightarrow s+t$. In any $d$, this symmetry is $s\rightarrow -s+2h-2, t\rightarrow s+t-h+1$. This symmetry has not been pointed out in the literature.
\end{itemize}
 Under $s\rightarrow -s+2h-2, t\rightarrow -t-h+1$ the polynomials acquire $(-1)^\ell$. This is just a combination of the first two transformations. The continuous Hahn polynomials respect $T_1$ but not $T_2$. For $\D-\ell=2(h-1)$ the Mack polynomials simplify dramatically. First the ${}_4F_3$ can be replaced by 1. Next we can just do the $m,n$ sums obtaining
 \be
  \widehat P^{(s)}_{\D-h,\ell}(s,t)=\frac{2^{-1-\ell}(s)_\ell\G(h+\ell-2)\G(2h+\ell-3)}{\G(2(h+\ell-2))\G(h-1)}{}_3F_{2}\left[-\ell,2h+\ell-3,-t;h-1,s;1\right]\,.
  \ee
 
\section{Spectral integral: $\ell=2$ case worked out}\label{C}

In this section we will explicitly work out the spectral integral for the $\ell=2$ case. This serves as an explicit example where the difference between the proposed exchange Witten diagram basis in this paper and in our earlier papers \cite{longpap, usprl} can be seen. We will focus on the $s$-channel part. We begin with
\be
\frac{1}{2\pi i}\int_{-i\infty}^{i\infty}d\nu \, \widehat P^{(s)}_{\nu,2}(s,t)\frac{\displaystyle\prod_{i=1,3}\Gamma(\frac{2\D_i+2-h}{2}-\frac{\nu}{2})\Gamma(\frac{2\D_i+2-h}{2}+\frac{\nu}{2})\Gamma(\frac{h-2-2s}{2}-\frac{\nu}{2})\Gamma(\frac{h-2-2s}{2}+\frac{\nu}{2})}{[(\Delta-h)^2-\nu^2]\Gamma(\D_1-s)\Gamma(\D_3-s)\Gamma(\nu)\Gamma(-\nu)}\,.
\ee
Here $\widehat P^{(s)}_{\nu,2}(s,t)$ is the spin-2 Mack polynomial which is related to our $P^{(s)}_{\nu,2}(s,t)$ via
$$
\widehat P^{(s)}_{\nu,\ell}(s,t)=\frac{ P^{(s)}_{\nu,\ell}(s,t)}{(h+\nu-1)_\ell (h-\nu-1)_\ell}
$$  
which leads a simple pole at $\nu=\pm(h-1)$ in $\widehat P^{(s)}_{\nu,\ell}(s,t)$.
We can write
\be
\frac{\widehat P^{(s)}_{\nu,\ell}(s,t)}{(\Delta-h)^2-\nu^2}=\frac{\widehat P^{(s)}_{\Delta-h,\ell}(s,t)}{(\Delta-h)^2-\nu^2} + F_{up}(s)+\tilde{P}_c\,,
\ee
where $F_{up}(s)$ is a polynomial in $s$ of degree 2 given by
$$
F_{up}(s)=\frac{(2s-2h+1)(2s-2h+3)}{128 h(\D-1)(\D-2h+1)}\,,
$$
and
$\tilde{P}_c=-\frac{1}{128h}$ is a constant. The subscript $up$ indicates unphysical; the reason for this nomenclature will become clear below. Now  the integrand for each piece above is of the form in (\ref{id1}). For all the three pieces we have
\be 
2a_1=2\D_1+2-h\,,\quad 2 a_3=2\D_3+2-h\,,\quad 2a_2=h-2-2s\,,
\ee
while $2a_4=\D-h$ for the first piece, $2a_4=h-1$ for the second and for the third we use \cite{AAR}
\be \label{id2}
\frac{1}{2\pi i}\int_{-i\infty}^{i\infty} \ d\nu \frac{\displaystyle\prod_{i=1}^3\Gamma(a_i-\frac{\nu}{2})\Gamma(a_i+\frac{\nu}{2})}{\Gamma(\nu)\Gamma(-\nu)}=4\Gamma(a_1+a_2)\Gamma(a_2+a_3)\Gamma(a_1+a_3)\,.
\ee
 We will now specialize to the identical scalar case where $\D_i=\D_\phi$. 
Explicitly we find
\begin{eqnarray}
&&-\frac{2 \widehat P^{(s)}_{\D+h,\ell}(s,t) \Gamma^2 \left(1-h+\frac{\Delta }{2}+\Delta_\phi \right)\, _3F_2\left[\frac{\Delta }{2}-s-1,\frac{\Delta }{2}-\Delta_\phi ,\frac{\Delta }{2}-\Delta_\phi ;\frac{\Delta }{2}-s,\Delta-h +1;1\right]}{(2+2s-\Delta) \Gamma (\Delta-h +1)}\nonumber \\&&+\frac{2F_{up}(s) \Gamma^2 \left(\Delta_\phi +\frac{1}{2}\right)  \, _3F_2\left[h-s-\frac{3}{2},h-\Delta_\phi -\frac{1}{2},h-\Delta_\phi -\frac{1}{2};h,h-s-\frac{1}{2};1\right]}{(2 h-2 s-3) \Gamma (h)}\nonumber \\&&+4 \tilde{P}_c \Gamma (2-h+2 \Delta_\phi)\,.
\end{eqnarray}
Now notice that the term proportional to $f_{up}(s)$ has poles in $s$ that do not correspond to physical  states or their descendants. Hence the nomenclature introduced earlier. If we write the $_3F_2$ in terms of Pochhammers, it turns out that part of this answer does not have any unphysical pole. Including this part and directly setting $s=\D_\phi$, the result is quite simple:
\be
-\frac{ \left(\Delta_\phi ^2 (\Delta_\phi +1)+(4 \Delta_\phi +2) t^2+2 \Delta_\phi  (2 \Delta_ \phi +1) t\right) \Gamma (2 \Delta_\phi +2-h)}{2(2 \Delta_\phi +1) (2-\Delta +2 \Delta_\phi ) (\Delta +2 \Delta_\phi -2 h+2)}\,,
\ee
and can be written in terms of the continuous Hahn polynomials in the normalization of our other papers as 
\be
\frac{ \Gamma (2 \Delta_\phi +2-h)}{4(\Delta -2 \Delta_\phi +2) (\Delta +2 \Delta_\phi -2 h+2)} Q^{2\D_\phi+2}_{2,0}(t)\,,
\ee
which agrees exactly with what we obtained in our other papers \cite{usprl, longpap}. Now in this paper, we will work only with the meromorphic part which is the piece proportional to $f_p(s,t)$ above. Then the difference between our new basis and the old one used in \cite{usprl, longpap} is explicitly given by 
\be
\Delta M^{(s)}_{\ell=2}=2F_{up}(s)\frac{\G^2(\Dphi+\frac{1}{2})}{(2s+3-2h)\G(h)}\left(1+\frac{(2s+3-2h)(2\Dphi-h+1)^2}{4h(2s+1-2h)}\right)-4\tilde{P}_c \Gamma (2-h+2 \Delta_\phi)\,.
\ee
Notice that $F_{up}(s)$ has zeros at $s=h-3/2$ and $s=h-1/2$ so that the difference is just a degree-2 polynomial in $s$. In the epsilon expansion this difference will show up at $O(\e^4)$ in the $\ell'\neq 0$ equations and at $O(\e^3)$ for the $\ell'=0$ equation. We further note that the split representation form of the propagator can be used to give different forms for the exchange diagram which differ by polynomial terms \cite{spinningads, charlotte}.

\section{Explicit Expressions: $t$-Channel Exchange}\label{appd}

\subsection{Scalar Exchange with $\ell'=0$}\label{ck}

In Secs. 3.2 and 3.3 we had decomposed the $t$-channel Witten diagram in terms of $s$-channel partial waves.  The expression we needed to evaluate was given in eq.(\ref{qcoeff3}). We had expressed this in terms of a potentially simpler integral. Thus for the simplest case of a scalar exchange the coefficient of the decomposition in the $\ell'=0$ partial wave we had the form given in eq.(\ref{qcoeff2}). In this appendix we first give various alternate forms of this integral and give exact expressions for certain simple cases. In the next subsection, we carry out the integral explicitly and simplify the expression into a finite sum of products of hypergeometric functions. 
These expressions are the analogue for Witten diagrams of what was recently carried out for the crossing kernel for the decomposition of $t$-channel conformal partial waves in the $s$-channel \cite{tar-slght, 6jdsd}. 

An example of an equivalent form is obtained by starting with an alternate form of the $t$-channel exchange diagram given by   
\begin{eqnarray}\label{texp}
M^{(t)}_{\D, \ell=0}(s,t) && = -\frac{2 \Gamma (2 \Delta_\phi -h) \Gamma \left(\frac{\Delta }{2}+\Delta_\phi -h\right)^2 \Gamma \left(-t+\frac{\Delta }{2}-\Delta_\phi +1\right) \, }{\Gamma (\Delta-h +1)\Gamma \left(-h-t+\frac{\Delta }{2}+\Delta_\phi \right)  (2 \Delta_\phi-\D +2 t) }\\
&&\!\!\!\!\!\!\times _3F_2\left[-t+\frac{\Delta }{2}-\Delta_\phi ,-h+\frac{\Delta }{2}+\Delta_\phi ,-h+\frac{\Delta }{2}+\Delta_\phi ;-h+\Delta +1,-h-t+\frac{\Delta }{2}+\Delta_\phi ;1\right]\,.  \nonumber
\end{eqnarray}
This is obtained by making a suitable transformation of the form Eq.(\ref{3f2t}) on the $s$-channel exchange diagram and then going to the $t$-channel by exchange of $(s,t)$. 

Using the general  integral representation of ${}_3F_2$:
%(cf. Eq.(\ref{3f2int1}) )
\be\label{3f2int}
{}_3F_2[a_0,a_1,a_2;b_0,b_1;1]=\frac{\G(b_0)}{\G(a_0)\G(b_0-a_0)}\int_0^1 dy\ y^{a_0-1}(1-y)^{b_0-a_0-1}{}_2F_1[a_1,a_2,b_1;y]\,,
\ee
where
\be
a_0=-t+\frac{\Delta }{2}-\Delta_\phi \,, \quad b_0=-h-t+\frac{\Delta }{2}+\Delta_\phi\,.
\ee
we can retrace the steps that follow such as doing the $t$-integral as in  Eq.(\ref{tintsimp}) to get the equivalent form
\begin{eqnarray}\label{qcoeffdi}
q^{(2,t)}_{\D, \ell'=0 |\ell=0} &=&\frac{\G^2(\frac{\D}{2}+\Dphi-h)}{\G(\D-h+1)}  \int_{0}^{1}  dy \,
 (1-y)^{2\Dphi-h-1}y^{\frac{\D}{2}-1}\\
& \times &{}_2F_1[\frac{\D}{2}+\Dphi-h,\frac{\D}{2}+\Dphi-h,\D-h+1;y]{}_2F_1[\Dphi,\Dphi,2\Dphi;1-y]  \nonumber\, .
\end{eqnarray}

Finally, starting with eq.(\ref{qcoeffdi}) we can also reach another useful form. We can expand the $2F_1[\frac{\D}{2}+\Dphi-h,\frac{\D}{2}+\Dphi-h,\D-h+1;y]$ around $y=0$ and ${}_2F_1[\Dphi,\Dphi,2\Dphi;1-y]$ around $y=1$ and carry out the integral over $y$. This gives an analytic continuation in the parameters and leads to
\be\label{tu0}
q^{(2,t)}_{\D, \ell'=0 |\ell=0}=\sum_{q,r=0}^{\infty}\frac{\G(2s)\G(s+r)^2\G(q+s+\frac{\D}{2}-\D_\phi)\G(q-h+\frac{\D}{2}+\D_\phi)^2\G(r+2\D_\phi-h)}{q!r!\G(s)^2\G(r+2s)\G(\D-h+1+q)\G(s+q+r+\frac{\D}{2}+\D_\phi-h)}{\bigg |}_{s=\Dphi}\,.
\ee 
For many purposes, especially for the $\e$-expansion, this form will be the most convenient one. 
 
We can evaluate the $y$-integrals using one of the forms given Eq.(\ref{qcoeff2}) or Eq.(\ref{qcoeffdi}  easily in some special cases. 

  {\bf A. $\Dphi=h/2$:} The $y$-integral in the form Eq.(\ref{qcoeff2})  is easily doable
\be
\frac{\G(h)\G^2(\frac{h}{2}+a-1)}{\G^2(\frac{h}{2})\G(a)\G(h+a-1)}\int_{0}^{1}  dy 
 y^{\frac{h}{2}-1}(1-y)^{a-2}{}_2F_1[\frac{h}{2},\frac{h}{2}+a-1;h+a-1,1-y]=\frac{1}{a-1} \,,
 \ee
 with $a=1+\frac{\D}{2}-\frac{h}{2}$ and we have used eq.(\ref{3f2int}) along with the Gauss summation formula to obtain the final step.
For this case, in the expression for  $q^{(2,t)}_{\D, \ell'=0 |\ell=0} $, the pre-factor $\G(2\Dphi-h)$ gives rise to a simple pole so separating this factor out we write 
\be
q^{(2,t)}_{\D, \ell'=0 |\ell=0} =\frac{4\G(2\Dphi-h)}{(\D-h)^2}\,.
\ee

 {\bf B. $\Dphi =1$:} In this case we start with Eq.(\ref{qcoeffdi} 
\begin{eqnarray}\label{qcoeffdi2}
q^{(2,t)}_{\D, \ell'=0 |\ell=0} &=&\frac{\G^2(\frac{\D}{2}+\Dphi-h)}{\G(\D-h+1)}  \int_{0}^{1}  dy \,
 (1-y)^{2\Dphi-h-1}y^{\frac{\D}{2}-1}\\
& \times &{}_2F_1[\frac{\D}{2}+\Dphi-h,\frac{\D}{2}+\Dphi-h,\D-h+1;y]{}_2F_1[\Dphi,\Dphi,2\Dphi;1-y]  \nonumber\, .
\end{eqnarray}
Then setting $\Dphi=1$ and using ${}_2F_1[1,1;2,1-y] =(y-1)^{-1}\log y$, we have
the following $y$-integral
\be
 \int_{0}^{1}  dy \,
 (1-y)^{1-h} y^{\frac{\D}{2}-1}
{}_2F_1[\frac{\D}{2}+1-h,\frac{\D}{2}+1-h,\D-h+1;y]\frac{\log y}{y-1}\,.
\ee
While convergence requires $Re[h]<2$ (which is certainly met in the $\e$-expansion if $\e>0$), we can analytically continue the final expression to other non-integral values of $h>2$. Expanding the ${}_2F_1$ and carrying out the $y$-integral, term by term, leads to
\be
\G(h-1)\sum_{n=0}^\infty \frac{\G(\frac{\D}{2}+n)\G(\frac{\D}{2}-h+1+n)}{n!\G(\D-h+1+n)}\left(\psi(\frac{\D}{2}+n-h+1)-\psi(\frac{\D}{2}+n)\right)\,.
\ee
To evaluate the sum at this stage, we write $\psi(a)=\partial_x \frac{\G(a+x)}{\G(a)}|_{x=0}$. Using this, the sum over $n$ can be carried out leading to
\be\label{exactck1}
q^{(2,t)}_{\D, \ell'=0 |\ell=0}=\G(1-h)\left(\psi'(\frac{\D}{2})-\psi'(\frac{\D}{2}-h+1)\right)\,.
\ee
Here $\psi'(x)$ is the tri-gamma function. In the special case $h=1,\Dphi=1$, we obtain $q^{(2,t)}_{\D, \ell'=0 |\ell=0}=\psi''(\frac{\D}{2})$.

\subsection{General Spin Exchange}
For the general spin exchange, after some work, we can give explicit expressions in terms of a single ${}_7F_6$! This remarkable result can be proven as follows. We start with the $y$ integral in eq.(\ref{qcoefft2s}):
\be
\int_{0}^{1}  dy 
 y^{s-1}(1-y)^{a_\ell-1}  {}_2F_1[1, a_\ell; a_\ell+\ell+(2\Delta_{\phi}-h);y]{}_2F_1[s+m+a_\ell-1,s+p;2s+p+m+a_\ell-1,1-y] 
\ee
 which after a Pfaff transformation i.e. use ${}_2F_1 (a, b; c; y) =(1-y)^{-a}\, {}_2F_1 (a, c-b; c ;\frac{y}{y-1})$, becomes
 \begin{eqnarray}
 \int_{0}^{1}  dy \, y^{-a_\ell-m}(1-y)^{a_\ell-2}&\times&{}_2F_1[s+m+a_\ell-1,s+m+a_\ell-1,2s+m+p+a_\ell-1,\frac{y-1}{y}]\nonumber\\ &\times& {}_2F_1[1,2\Dphi-h+\ell,2\Dphi-h+\ell+a_\ell,\frac{y}{y-1}]\,.
\end{eqnarray}
Next incorporating the $p$-dependent factors from eq.(\ref{qcoefft2s}) and carrying out the sum over $p$ (using the Chu-Vandermonde identity or the Gauss summation formula at the last step) we find
\begin{eqnarray}\label{tchangen}
&I=&\int_0^1 dy\, y^{-a_\ell-m}(1-y)^{a_\ell-2}\G(a_\ell+m){}_2F_1[1,2\Dphi-h+\ell,2\Dphi-h+\ell+a_\ell,\frac{y}{y-1}]\nonumber \\ &\times& {}_3\tilde F_2[a_\ell+m, a_\ell+m+s-1, a_\ell+m+s-1; a_\ell+m-\ell',a_\ell+m+\ell'+2s-1;\frac{y-1}{y}]\,.\nonumber\\
\end{eqnarray}
Here ${}_p\tilde{F}_q$ is the regularized version of the hypergeometric function\footnote{Defined by ${}_p\tilde{F}_q(\{a_i\};\{b_i\};z) = \frac{1}{\prod_{i=1}^q\G(b_i)} \, {}_pF_q(\{a_i\};\{b_i\};z).$}. After this we use the Mellin-Barnes representation
\be
{}_pF_q[\{a_i\};\{ b_k\}; x]=\frac{\prod_{k}\G(b_k)}{\prod_i \G(a_i)}\int_{\gamma-i\infty}^{\gamma+i\infty}d\sigma\, \frac{\prod_i \G(a_i+\s)}{\prod_k \G(b_k+\s)}\G(-\s)(-x)^\s\,,
\ee
to obtain the following form for the $y$-integral ($\sigma$ is the MB variable for the ${}_3F_2$ and $\tau$ is the MB variable for the ${}_2F_1$)
\be
\int_{\g_1-i\infty}^{\g_1+i\infty}d\tau\, \int_{\g_2-i\infty}^{\g_2+i\infty} d\s\,F(\sigma,\tau)\int_0^1 \frac{dy}{(1-y)^2}(\frac{y}{1-y})^{\tau-\s-a_\ell}y^{-m}\,.
\ee
At this stage, we make the change of variables $y/(1-y)=z$ to get the integral
\be
\int_{\g_1-i\infty}^{\g_1+i\infty}d\tau\, \int_{\g_2-i\infty}^{\g_2+i\infty}d\s\, F(\sigma,\tau)\int_0^\infty dz \, z^{-a_\ell-m-\s+\tau}(1+z)^m\,.
\ee
Now we can write the resulting $(1+z)^m=\sum_{k=0}^m {}^m C_k z^k$ followed by the change of variables $z=e^{x}$.  Then we choose $\g_1/\g_2$ to make $Re[\tau-\s-m-a_\ell+k+1]=0$ along the contour. This leads to 
\be
2\pi i \sum_{k=0}^m {}^m C_k \int_{\g_1-i\infty}^{\g_1+i\infty}d\tau\, \int_{\g_2-i\infty}^{\g_2+i\infty} d\s\, F(\sigma,\tau) \delta(\tau-\s-a_\ell+k-m+1)\,.
\ee
This enables us to do the $\tau$ integral straightaway and allows us to do the $k$-sum. The final expression is
\begin{eqnarray}
I&=&(-1)^{\ell'+m}\int_{\g_2-i\infty}^{\g_2+i\infty} d\s \frac{(a_\ell)_m\G(2\Dphi-h+\ell+a_\ell)}{\G^2(a_\ell+s+m-1)\G(2\Dphi-h+\ell)}\nonumber\\
&\times&\frac{\G(-\s)\G(1-a_\ell+\ell'-m-\s)\G(a_\ell+m+\s)\G^2(a_\ell+m+s-1+\s)\G(2\Dphi-h+\ell-1+a_\ell+\s)}{\G(a_\ell+\ell'+m+2s-1+\s)\G(2a_\ell+\ell+m+2\Dphi-h-1+\s)}\,.\nonumber\\
\end{eqnarray}
Now quite remarkably, the $\s$ integral is precisely of the form for a very well-poised ${}_7F_6$ at unit argument which seems to have amazing applications in mathematics. See for example eq.(2) in \cite{161108806} which is 
\begin{eqnarray}\label{Wdef}
&&W(a;b,c,d,e,f)\equiv \nonumber\\
&&{}_7F_6\bigg{(}\begin{matrix} a, & 1+\frac{1}{2}a, & b, & c, & d, & e, & f\\ ~&\frac{1}{2}a,& 1+a-b, & 1+a-c, & 1+a-d, & 1+a-e, & 1+a-f\end{matrix};1\bigg{)}\nonumber\\
&=&\frac{\G(1+a-b)\G(1+a-c)\G(1+a-d)\G(1+a-e)\G(1+a-f)}{\G(1+a)\G(b)\G(c)\G(d)\G(1+a-c-d)\G(1+a-b-d)\G(1+a-b-c)\G(1+a-e-f)}\nonumber\\
&\times&\frac{1}{2\pi i}\int_{-i\infty}^{i\infty}d\s\, \frac{\G(-\s)\G(1+a-b-c-d-\s)\G(b+\s)\G(c+\s)\G(d+\s)\G(1+a-e-f+\s)}{\G(1+a-e+\s)\G(1+a-f+\s)}\,.\nonumber\\
\end{eqnarray}
The $W$ notation was introduced by Bailey \cite{bailey} and we will use it to shorten our expressions--related expressions in terms of ${}_4F_3$'s appear in the 1-d bootstrap through what are called Wilson functions \cite{vanrees}.
For our case we have 
\be\label{params}
a=\ell'+2(a_\ell+m+s-1)\,,\quad b=e=a_\ell+m,\quad c=d=a_\ell+m+s-1\,,\quad f=2(s-\Dphi)+h+m+\ell'-\ell\,.
\ee
For the above expression to be finite, we need $4a-2(b+c+d+e+f-2)>0$. When this is not satisfied\footnote{Depending on the sign of certain combination of parameters, we may need to add a finite number of extra terms. This is not needed for the $\ell'=\ell=0$ case in the epsilon expansion.}, we need to analytically continue the expression (for instance in the epsilon expansion for non-zero spins in the t-channel, this condition is not respected for $m=\ell$ and we work with the original $y$ integral form; alternatively we could have worked out an analytic continuation but we will not do so here).
 These functions have been studied in the mathematics literature and are the generalizations of the $su[2], su[1,1]$ $6j$ symbols \cite{6jgen}. Furthermore, they can also be written as the difference of two well balanced ${}_4F_3$ hypergeometric functions--to derive this form one can simply use the InverseMellinTransform command in mathematica using the $\s$-integrand above. Recently such a form in terms of the difference between ${}_4F_3$'s have been reported as the $6j$-symbols appearing in the crossing kernel of the usual conformal bootstrap formalism \cite{6jdsd} but the connection with ${}_7 F_6$ was not realized. For the equal scalar case (which is the case of interest in our paper), this difference in fact (see eq.(3.36) which is for 2 spacetime dimensions in \cite{6jdsd,vanrees} where in our notation $a=2h'+h-1, b=c=h',d=1+h'-p,e=h, f=h'+p-1$) is again a single well-poised ${}_7F_6$. A final observation is that for $\ell'=0$ the ${}_7F_6$ becomes a ${}_5F_4$.
 
The explicit final form is:
\be
q^{(t)}_{\D,\ell'|\ell}(s)=\sum_{m,n}^{\ell} (-1)^{\ell'+m}2^{-\ell'}\mu_{m,n}^{(\ell)} (\Dphi-s)_n (a_\ell)_m^2 \G(2s+2\ell')\G^2(d)\G(\frac{a}{2})\G(a+1)\G^2(1+a-f-b)\tilde W(a;b,c,d,e,f)\,,
\ee
where the parameters are given in eq.(\ref{params}) and the $\tilde W$ is the regularized version of $W$. We can explicitly cross-check this expression for various limiting cases where exact expressions are known (eg. eq.(\ref{exactck1})). We will come back to a detailed study of this remarkable final form in future work.

\end{document}